\documentstyle[latexsym,amssymb,epsfig,aps,floats,preprint,eqsecnum,amsfonts]{revtex}

\def\be{\begin{equation}}
\def\ee{\end{equation}}

\begin{document}

\title{The entropy of black holes: a primer}
\author{Thibault Damour}
\address{{\sl Institut des Hautes Etudes Scientifiques, 35 route de Chartres, 91440 
Bures-sur-Yvette, France}}
\maketitle

\begin{abstract}
After recalling the definition of black holes, and reviewing their energetics and
their classical thermodynamics, one expounds the conjecture of Bekenstein,
attributing an entropy to black holes, and the calculation by Hawking of the
semi-classical radiation spectrum of a black hole, involving
a thermal (Planckian) factor.  One then discusses the attempts to interpret the black-hole
entropy as the logarithm of the number of quantum micro-states of a
macroscopic black hole, with particular emphasis on results obtained within
string theory.  After mentioning the (technically cleaner, but conceptually
more intricate) case of supersymmetric (BPS) black holes and the corresponding
counting of the degeneracy of  Dirichlet-brane systems, one discusses
in some detail the ``correspondence'' between massive string states
and non-supersymmetric Schwarzschild black holes. 

\end{abstract}

\section{Black holes}

Let us start by briefly reviewing the concept of black hole. Within a couple of 
months after Einstein's discovery of the final form of the field equations of General 
Relativity, Karl Schwarzschild succeeded in writing down the exact form of the 
general relativistic analog of the ``gravitational field of a mass point'' 
\cite{S16}, namely (in the coordinates introduced by Johannes Droste) the general 
spherically-symmetric solution of Einstein's vacuum field equations $(R_{\mu\nu} = 
0)$:
\be
\label{eq1.1}
ds_{\rm SCHW}^2 = g_{\mu\nu}^{\rm SCHW} (x^{\lambda}) \, dx^{\mu} \, dx^{\nu} = - 
\left( 1 - \frac{2 GM}{c^2 r} \right) c^2 \, dt^2 + \frac{dr^2}{1 - \frac{2 GM}{c^2 
r}} + r^2 (d\theta^2 + \sin^2 \, \theta \, d \varphi^2) \, .
\ee
Here, $G$ denotes Newton's constant and $M$ denotes the (total) mass of the 
considered object, as it can be measured from infinity ({\it e.g.} by comparing the 
motion of far-away test particles in the metric (\ref{eq1.1}) to that of test 
particles in the Newtonian potential $U ({\mathbf x}) = GM/r$).

It was soon noticed that the metric (\ref{eq1.1}) has an apparently singular 
behaviour at the ``Schwarzschild radius'' $r_S \equiv 2GM/c^2$. For instance, a 
clock at rest in the metric (\ref{eq1.1}) and located at a radius $r$ ($r > r_S$) 
exhibits, when its ticks are 
 ``read'' from infinity via electromagnetic signals, a redshift equal to
\be
\label{eq1.2}
\frac{[ds]^{\text{read at}}_{\text{infinity}}} 
{[ds]^{\text{measured}}_{\text{locally}}} = \frac{\sqrt{-g_{00} 
(r=\infty)}}{\sqrt{-g_{00} (r)}} = \frac{1}{\sqrt{1 - \frac{2GM}{c^2 r}}} \, .
\ee

The redshift (\ref{eq1.2}) goes to infinity as $r \to r_S$ . Therefore the 
``Schwarzschild sphere'' $r = r_S = 2GM/c^2$ has the observable characteristic of 
being an infinite-redshift surface.  Similarly, one finds that the force needed to
keep a particle at rest at a radius $r > r_S$ goes to infinity as $r \to r_S$.

\begin{figure}[ht]
\begin{center}
\includegraphics[scale=.8]{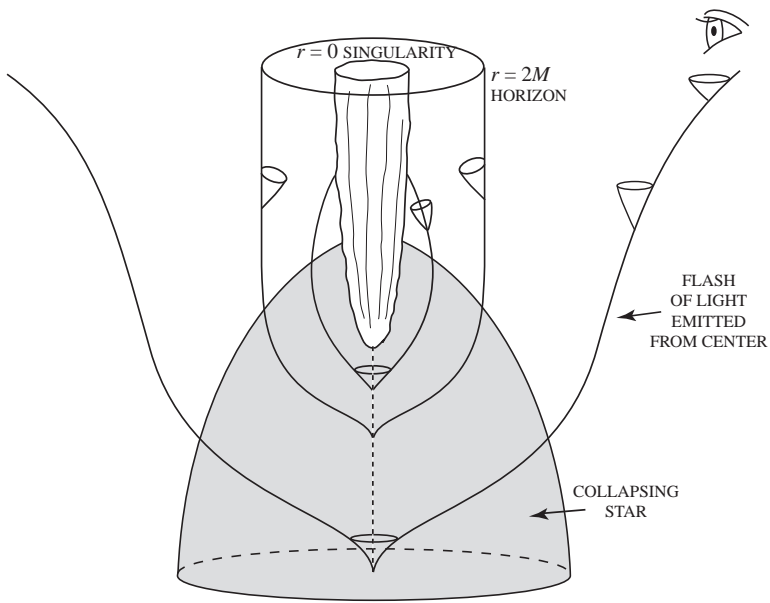}
\vglue 5mm
\caption{Spacetime representation of the formation of a black hole by the
collapse of a star. The horizon is the spacetime history of a bubble
of light ({\it i.e.} a null hypersurface)
which stabilizes itself under the strong pull of  relativistic gravity}
\label{fig1}
\end{center}
\end{figure}

It took many years, and the work of many 
scientists (notably Oppenheimer and Snyder, Kruskal and Penrose), to decipher the 
physical meaning of these infinities occurring at the
 ``Schwarzschild sphere''. This understanding is summarized 
in the spacetime diagram of Fig.~\ref{fig1}.

\noindent The 2-dimensional surface $r = r_S$, becomes, when adding the time 
dimension, a 3-dimensional hypersurface ${\mathcal H}$ in spacetime. The hypersurface 
${\mathcal H}$ is a fully regular submanifold of a locally regular spacetime. 
To see the regularity of ${\mathcal H}$, one needs to change coordinates near $r=r_S$.
For instance, one can use the ingoing Eddington-Finkelstein coordinates
$(v,r,\theta,\varphi)$, where $v=t+r_{*}$, with $r_*$ defined as
\be
\label{eq2.7pre}
r_* = \int \frac{dr}{1 - \frac{2GM}{c^2r}} = r + \frac{2GM}{c^2} \ln \left( 
\frac{c^2r}{2GM}-1 
\right) \, 
.
\ee
In these coordinates the Schwarzschild metric reads
\be
\label{EF}
ds_{\rm SCHW}^2 = 
- \left( 1 - \frac{2 GM}{c^2 r} \right) \, dv^2 + 2 dv dr 
+ r^2 (d\theta^2 + \sin^2 \, \theta \, d \varphi^2) \, .
\ee
In these coordinates the horizon is located at $r=r_S$, the other coordinates
$(v,\theta,\varphi)$ taking arbitrary, but finite values. [Note that a finite value
of the new ``time variable'' $v$ corresponds to an infinite (and positive)
value of the original Schwarzschild time coordinate $t$.] One easily checks that
the geometry ( \ref{EF}) is regular near $r=r_S$.

One can then see that the hypersurface  ${\mathcal H}$ ($r=r_S$) is a quite
special submanifold in that it is  a {\it null hypersurface}, 
{\it i.e.} a co-dimension-1 surface which is locally tangent to the light cone. In 
other words, the (co-)vector normal to the hypersurface, say $\ell_{\mu}$ (such that 
$\ell_{\mu} \, dx^{\mu} = 0$ for all directions $dx^{\mu}$ tangent to the 
hypersurface) is a null vector: $0 = g^{\mu\nu} \ell_{\mu} \, \ell_{\nu} = g_{\mu\nu} \, 
\ell^{\mu} \ell^{\nu}$. As a consequence $\ell^{\mu}$ is both normal and tangent to 
the hypersurface (because $\ell_{\mu} \, \ell^{\mu} = 0$). The local light cone, 
$g_{\mu\nu} (x) \, dx^{\mu} \, dx^{\nu} = 0$,  is tangent to ${\mathcal H}$ along the 
special direction $\ell^{\mu}$. [This gives a special, fibered structure to 
${\mathcal H}$, generated by the lines tangent to $\ell^{\mu}$, called the 
``generators'' of ${\mathcal H}$.] Physically, the tangency between ${\mathcal H}$ 
and the local light cone (and the fact that the spatial sections of ${\mathcal H}$ are 
compact) means that ${\mathcal H}$ is the boundary between the part of spacetime from 
which light can escape to infinity, and the part out of which light cannot so escape. 
This boundary is called the (future) {\it Horizon} (hence the notation ${\mathcal 
H}$).

Fig.~\ref{fig1} illustrates the fact that the Horizon ${\mathcal H}$ is a dynamical, 
time-evolving structure. In the simple, spherically-symmetric situation assumed in 
Fig.~\ref{fig1}, the horizon ${\mathcal H}$ is the spacetime history of a special 
bubble of light that was emitted from the center of a collapsing star, and which, 
after expanding out from the center, stabilized itself, under the strong pull of 
relativistic gravity, into an asymptotically stationary configuration, which is 
precisely the (time-independent) Schwarzschild sphere $r = r_S \equiv 2GM/c^2$. The 
structure illustrated by Fig.~\ref{fig1} is called a {\it black hole}\footnote{
Note the interesting historical coincidence that the name of the discoverer of the
solution Eq.(\ref{eq1.1}) means  {\it black shield}.}. The 
``interior'' of ${\mathcal H}$ is the interior of the black hole (spacetime region 
where light is ``trapped'', {\it i.e.} cannot escape to infinity), while ${\mathcal 
H}$ (which marks the great ``divide'' between trapped light, and light escaping at 
infinity) is the horizon, or the {\it surface of the black hole}. Note the presence, 
inside the black hole, of an infinite spacetime volume which ends, in its future, in 
a spacelike singularity at $r=0$ (where the curvature blows up as $r^{-3}$). This 
singularity is a ``big crunch'' singularity: it is of a cosmological type (the 
coordinate $r$ of Eq.~(\ref{eq1.1}) being ``time-like'' when $r < r_S$), and a 
time-reverse analog of the familiar Friedmann big-bang singularity.

A lot of work (for references, see, {\it e.g.}, \cite{BH72,HE73,MTW,Wald,Heusler}) 
went into understanding how general is the picture of 
Fig.~\ref{fig1}. The stability (near and outside ${\mathcal H}$) of the spherical 
collapse situation of Fig.~\ref{fig1} under small, non-spherically-symmetric 
perturbations, the existence of multi-parameter generalizations of the Schwarzschild 
metric (due to Reissner, Nordstr\o m, Kerr, Newman {\it et al.}) featuring regular 
horizon structures, and the proof of several ``no-hair'' theorems (by Israel, Carter, 
etc.), led to the following (conjectural) picture. The gravitational collapse of any 
type of matter configuration will generically lead to the formation of a black hole, 
whose surface is a null hypersurface ${\mathcal H}$, and will asymptotically settle 
into a stationary state, which is completely described (if the black hole is 
isolated) by a small number of parameters. When considering, as long range fields, 
only gravity $(g_{\mu\nu})$ and electromagnetism $(A_{\mu})$, the final, stationary 
configuration of a black hole is described by three parameters, its total mass $M$, 
its total angular momentum $J$, and its total electric charge $Q$, and is given by 
the following Kerr-Newman solution
\begin{mathletters}
\label{eq1.3}
\begin{eqnarray}
ds_{\rm KN}^2 &= & - \frac{\Delta}{\Sigma} \, \omega_t^2 + \frac{\Sigma}{\Delta} \, 
dr^2 + \Sigma d\theta^2 + \frac{\sin^2 \theta}{\Sigma} \, \omega_{\varphi}^2 \, ,
\label{eq1.3a} \\
A_{\mu}^{\rm KN} \, dx^{\mu} &=& - \frac{Q r}{\Sigma} \omega_t \, ,
\label{eq1.3b} \\
\frac{1}{2} \, F_{\mu\nu}^{\rm KN} \, dx^{\mu} \wedge dx^{\nu} &= &\frac{Q}{\Sigma^2} 
\, (r^2 - a^2 \cos^2 \theta) \, dr \wedge \omega_t + \frac{2Q}{\Sigma^2} \, a r \cos 
\theta \sin \theta \, d \theta \wedge \omega_{\varphi} \, ,
\label{eq1.3c} 
\end{eqnarray}
\end{mathletters}
where (setting for simplicity $G = c = 1$)
$$
a \equiv \frac{J}{M} \, , \quad \Delta = r^2 - 2Mr + a^2 + Q^2 \, , \quad \Sigma = 
r^2 + a^2 \cos^2 \theta \, ,
$$
\be
\label{eq1.4}
\omega_t = dt - a \sin^2 \theta \, d\varphi \, , \quad \omega_{\varphi} = (r^2 + a^2) 
\, d\varphi - a \, dt \, .
\ee

The Kerr-Newman field configuration (\ref{eq1.3}) is a black hole (with a regular 
horizon ${\mathcal H}$) if and only if the three parameters $M$, $J$, $Q$ satisfy the 
inequality (recall the notation $a \equiv J/M$, and the choice of units $c = G = 1$
that will often be used below)
\be
\label{eq1.5}
a^2 + Q^2 \leq M^2 \, .
\ee
The black holes that saturate the inequality (\ref{eq1.5}), {\it i.e.} those that 
satisfy $a^2 + Q^2 = M^2$, are called {\it extremal}. Note that a Schwarzschild black 
hole ($a = Q = 0$) can never be extremal, while a Reissner-Nordstr\o m black hole 
($a=0$) is extremal when $\vert Q \vert = M$ ({\it i.e.} $\vert Q \vert = 
G^{\frac{1}{2}} M$), and a Kerr black hole ($Q=0$) is extremal when $\vert a \vert = 
M$ ({\it i.e.} $J = GM^2$). By analyzing the spacetime geometry (\ref{eq1.3a}) one 
finds that the horizon of a Kerr-Newman black hole is located (when viewed at 
``constant time'') on the surface\footnote{Contrary to the simple Schwarzschild case 
where the natural time sections of the horizon were metric spheres (with uniform 
Gauss curvature $K = r_S^{-2}$), the natural time sections of the Kerr-Newman horizon 
are {\it not} metric spheres, {\it i.e.} the Gauss curvature of their inner geometry 
is not uniform.} $r = r_+$ with
\be
\label{eq1.6}
r_+ \equiv M + \sqrt{M^2 - a^2 - Q^2} \, .
\ee

\section{Energetics and thermodynamics of black holes}

It is very fruitful to consider a black hole as a kind of {\it gravitational 
soliton}, {\it i.e.} as a physical object, localized ``within the horizon ${\mathcal 
H}$'', generating the ``external fields'' (\ref{eq1.3}), and possessing a total mass 
$M$, a total angular momentum $J$ and a total electric charge $Q$. By dropping 
massive, charged test particles, moving on generic non radial orbits, one then 
expects that it is possible to change the values of $M$, $J$ and $Q$, {\it i.e.} to 
evolve from an initial Kerr-Newman black hole state $(M,J,Q)$ to a final one 
($M+\delta M$, $J + \delta J$, $Q + \delta Q$) where $\delta M = E$, $\delta J = 
p_{\varphi}$ and $\delta Q = e$. Here $E = -p_t = - \partial S / \partial t$ is the 
conserved energy of the test particle freely moving (geodesic motion) in the 
background Kerr-Newman configuration $(M,J,Q)$, $p_{\varphi} = \partial S / \partial 
\varphi$ its conserved ($z$-component of the) angular momentum, and $e$ its electric 
charge. [$S$ denotes here the action of the test particle, satisfying the 
Hamilton-Jacobi equation $g^{\mu\nu} (p_{\mu} - e A_{\mu}) (p_{\nu} - e A_{\mu}) = 
-\mu^2$, where $p_{\mu} = \partial S / \partial x^{\mu}$ is its 4-momentum and $\mu$ 
its mass.] It was first realized by Penrose \cite{P69} that such a process can lead, 
in certain (special) circumstances, to a {\it decrease} of the total mass of the 
black hole: $\delta M < 0$. A detailed analysis of the changes of $M$, $J$ and $Q$ 
during the absorption of test particles, then led Christodoulou \cite{C70} and 
Christodoulou and Ruffini \cite{CR71} to introduce the concepts of {\it reversible 
transformation} and of {\it irreversible transformation} of a black hole. More 
precisely, starting from $\delta M = -p_t$, $\delta J = p_{\varphi}$, $\delta Q = e$ 
and from the mass-shell constraint $g^{\mu\nu} (p_{\mu} - e A_{\mu}) (p_{\nu} - e 
A_{\nu}) = -\mu^2$ (which gives a quadratic equation to determine the
energy  $ -p_t$ in terms of $p_{\varphi}$, $e$ and the other components of 
$p_{\mu}$, namely $p_r$ and $p_{\theta}$), they derived, when considering
a particle crossing the horizon, {\it i.e.} being absorbed by the black hole,
 the following equality 
\be
\label{eq1.7n}
\delta M  -  \frac{a \, \delta J + r_+ \, Q \, \delta Q}{r_+^2 + a^2}  =
\frac{r_+^2 + a^2 \cos^2 \theta}{r_+^2 +a^2} \vert p^r \vert \, ,
\ee
the right-hand side of which is proportional to the {\it absolute value}
of the radial momentum of the particle as it enters the black hole. 
[The latter absolute value comes from taking a positive square root,
similarly to the usual special relativistic result 
$E = + \sqrt{{\mu}^2 + {\bf p}^2}$.]
 The positivity of the
latter quantity then leads to deriving
 the {\it inequality}
\be
\label{eq1.7}
\delta M \geq \frac{a \, \delta J + r_+ \, Q \, \delta Q}{r_+^2 + a^2} \, .
\ee
A {\it reversible} transformation is then defined as a transformation which saturates 
the inequality (\ref{eq1.7}), {\it i.e.} where one replaces $\geq$ by an equality 
sign. It is called ``reversible'' because, after having (algebraically) added $\delta 
J$ and $\delta Q$ (and the corresponding $\delta M = $ R.H.S. of (\ref{eq1.7})) to a 
black hole, the subsequent reversible ``addition'' of $\delta' J = - \delta J$, 
$\delta' Q = -\delta Q$ (and the corresponding, saturated $\delta' M = -\delta M$) 
will lead to a final state $(M + \delta M + \delta' M, J + \delta J + \delta' J , Q + 
\delta Q + \delta' Q) \equiv (M,J,Q)$, {\it i.e.} identical to the initial state. By 
contrast, a sequence of infinitesimal transformations where at least one of the 
elementary processes (\ref{eq1.7}) contains the strict inequality sign $>$ cannot be 
reversed.

By integrating the differential equation obtained by saturating (\ref{eq1.7}) one 
obtains the Christodoulou-Ruffini {\it mass formula} of black holes,
\be
\label{eq1.8}
M^2 = \left( M_{\rm irr} + \frac{Q^2}{4 M_{\rm irr}} \right)^2 + \frac{J^2}{4 M_{\rm 
irr}^2} \, ,
\ee
where the {\it irreducible mass} $M_{\rm irr}$ is an (integration) {\it constant} 
under reversible processes, and varies in the following irreversible manner,
\be
\label{eq1.9}
\delta M_{\rm irr} \geq 0 \, ,
\ee
under general, possibly irreversible, processes. One also finds that the irreducible 
mass can be explicitly (rather than implicitly as in Eq.~(\ref{eq1.8})) expressed in 
terms of $M$, $J$ and $Q$ as
\be
\label{eq1.10}
M_{\rm irr} = \frac{1}{2} \, \sqrt{r_+^2 + a^2} \, .
\ee
Note that Eq.~(\ref{eq1.8}) says that the ``free energy'' of a black hole, {\it i.e.} 
the maximal energy which can be extracted by depleting (in a reversible) way $J$ and 
$Q$ is $M - M_{\rm irr}$. This free energy has both Coulomb $(\propto Q^2)$ and 
rotational $(\propto J^2)$ contributions. It vanishes in the case of a Schwarzschild 
black hole ($J = 0 = Q$).

For our present purpose, a crucial aspect of the global energetics of black holes, 
given by Eq.~(\ref{eq1.8}), is the striking similarity of the irreversible increase 
(\ref{eq1.9}) of the irreducible mass with the second law of thermodynamics, {\it 
i.e.} the irreversible increase of the total entropy $S$ of an isolated system. A 
general theorem of Hawking~\cite{H73} led to a deeper understanding of the 
irreversible behaviour (\ref{eq1.9}). Indeed, starting from the basic geometrical 
fact that a horizon is a null hypersurface, and analyzing the global properties of 
the generators of the horizon, Hawking proved that the area $A$ of successive time 
sections\footnote{By ``time sections'' of the horizon we mean some slices $v = cst$
where $v$ is some generalization of the (regular) Eddington-Finkelstein time coordinate.}
 of the horizon of a black hole cannot decrease,
\be
\label{eq1.11}
\delta A \geq 0 \, .
\ee
He also proved that, if one considers a system made of several, separate black holes, 
the sum of the areas of all the horizons cannot decrease
\be
\label{eq1.12}
\delta \left( \sum_a A_a \right) \geq 0 \, .
\ee
In the particular case of a Kerr-Newman hole the result (\ref{eq1.11}) yields back 
(\ref{eq1.9}). Indeed, one easily finds from Eq.~(\ref{eq1.3a}), when using $r = r_+$ 
({\it i.e.} $\Delta =0$ and $dr = 0$) that the inner geometry of the horizon is $d 
\sigma^2 = \gamma_{AB} (x^C) \, dx^A \, dx^B = (r_+^2 + a^2 \cos^2 \theta) \, d 
\theta^2 + \sin^2 \theta (r_+^2 + a^2)^2 \, d \varphi^2 / (r_+^2 + a^2 \cos^2 
\theta)$ (with $A,B = 1,2 = \theta , \varphi$ in the example at hand), so that the 
area of (a time section) of the horizon is
\be
\label{eq1.13}
A_{\rm KN} = \int \int (r_+^2 + a^2) \sin \theta \, d\theta \, d\varphi = 4\pi (r_+^2 + 
a^2) = 16\pi M_{\rm irr}^2 \, .
\ee

In view of Eqs.~(\ref{eq1.9}), (\ref{eq1.11}), (\ref{eq1.12}) it is tempting to 
attribute to a black hole, considered as a physical object, not only a total mass 
$M$, a total angular momentum, and a total electric charge, but also a total {\it 
entropy}, proportional to the area of the horizon, say
\be
\label{eq1.14}
S_{\rm BH} = \alpha A \, ,
\ee
where $\alpha$ is a constant with dimension of inverse length squared. By varying the 
mass formula (\ref{eq1.8}) one then finds the ``first law of the thermodynamics of 
black holes'',
\be
\label{eq1.15}
dM = \Omega \, d J + \Phi \, d Q + T_{\rm BH} \, d S_{\rm BH} \, ,
\ee
where (see Eq.~(\ref{eq1.7})) $\Omega = a / (r_+^2 + a^2)$ can be interpreted as the 
angular velocity of the black hole, $\Phi = Q \, r_+ / (r_+^2 + a^2)$ as its electric 
potential (see \cite{BH72} for discussions of $\Omega$ and $\Phi$), and where
\be
\label{eq1.16}
T_{\rm BH}  \equiv  \frac{1}{\alpha} \frac{\partial M}{\partial A} = 
\frac{\kappa}{8\pi\alpha}
\ee
is expected to represent the ``temperature'' of the black hole. The quantity $\kappa$ 
in Eq.~(\ref{eq1.16}) is the {\it surface gravity} of a black hole, generally defined 
as the coefficient relating  the covariant directional derivative of the horizon normal 
vector $\ell^{\mu}$ along itself to $\ell^{\mu} : \ell^{\nu} \, \nabla_{\nu} \, 
\ell^{\mu} = \kappa \, \ell^{\mu}$. [Here, $\ell^{\mu}$ is normalized so that, on the 
horizon, $\ell^{\mu} \, \partial_{\mu} = \partial_t + \Omega \, \partial_{\varphi}$ 
where $\partial_t$ is the usual Killing vector of time translations.] The surface gravity
may be thought of as the {\it redshifted} acceleration of a particle staying ``still'' on 
the horizon.
[As we said above, the proper-time acceleration of a particle sitting on the horizon
is actually infinite, but the infinite redshift factor associated to the difference
between the proper time and the ``time'' associated to the generator  $\ell^{\mu}$
compensates for this infinity.]

In the case of a Kerr-Newman hole, the value of the surface gravity reads $(r_{\pm} \equiv
M \pm \sqrt{M^2 - a^2 - Q^2})$
\be
\label{eq1.17}
\kappa = \frac{1}{2} \, \frac{r_+ - r_-}{r_+^2 + a^2} = \frac{\sqrt{M^2 - a^2 - 
Q^2}}{r_+^2 + a^2} \, .
\ee
Note that the surface gravity of a Schwarzschild black hole is given by the usual 
formula for the surface gravity of a massive star, {\it i.e.} $\kappa = GM / r_S^2 = 
M / (2M)^2 = 1/(4M)$. Note also that the surface gravity (and therefore the expected 
``temperature'') of extremal black holes {\it vanishes}.

The ``first law of black-hole thermodynamics'' (\ref{eq1.15}) has been derived above 
in the parti\-cular (but particularly important!) case of Kerr-Newman black holes. A 
general derivation, for generic (non isolated) black-hole equilibrium states, has 
been given by Carter \cite{C73} and by Bardeen, Carter and Hawking \cite{BCH73}. 
For a recent review of black-hole thermodynamics, see \cite{Wald01}. See 
also these references for a discussion of the ``zeroth law of black-hole 
thermodynamics'', namely the constancy of the ``temperature'' (\ref{eq1.16}), {\it 
i.e.} of the surface gravity $\kappa$, all over the horizon. [In the ``membrane'' 
approach to black-hole physics \cite{D79}, the uniformity of $\kappa$ for stationary 
states is viewed as a consequence of the Navier-Stokes equation because the ``surface 
pressure'' of a black hole happens to be equal to $p \equiv \kappa / 8\pi$.]

Before going on to considering {\it quantum} aspects of the irreversibility of black 
hole evolution, let us complete our brief description of the {\it classical} dynamics 
and thermodynamics of black holes by mentioning that the irreversibility 
(\ref{eq1.9}) or (\ref{eq1.11}) affecting a {\it global} characteristic of a black 
hole has its roots in the irreversible behaviour of a {\it local} characteristic of 
the horizon. It proved fruitful to view a black hole as a kind of {\it membrane} 
\cite{D78,Z78,D79}, fibered by the generators (see above), and endowed with familiar 
physical properties: pressure $p = \kappa / 8\pi$, shear viscosity $\eta$, bulk viscosity 
$\zeta$, 
resistivity $\rho$. For instance, one can write an Ohm's law relating the 
electromotive force tangent to the horizon to the current flowing in the horizon, and 
a Navier-Stokes equation relating the time-evolution of the black-hole surface 
momentum density to the (surface) gradient of the pressure, to the effects of the 
shear and of the expansion of the ``2-dimensional'' fluid of generators and to a  
possible external flux of momentum. Starting from Einstein's equations one finds, 
{\it e.g.}, that the (surface) resistivity of black holes is $\rho = 377$ ohm ({\it 
i.e.} $4\pi$), and that their (surface) shear viscosity is $\eta = (16\pi)^{-1}$. Of 
particular interest for our present purpose is the existence of a {\it local} ({\it 
i.e.} at each point of the surface of the black hole) version of the ``second law of 
thermodynamics''. Let us ``localize'' the global attribution (\ref{eq1.14}) by 
attributing an {\it entropy}
\be
\label{eq1.18}
d S_{\rm BH} = \alpha \, dA
\ee
to each area element $dA$ of the horizon. This local entropy is then found to evolve, 
as a consequence of Einstein's equations, in the following irreversible manner 
\cite{HH72,D79}
\be
\label{eq1.19}
\frac{d}{dt} \, (dS_{\rm BH}) - \tau \, \frac{d^2}{dt^2} \, (dS_{\rm BH}) = 
\frac{d A}{T_{\rm BH}} \left[2 \eta \, \sigma_{\rm AB} \, \sigma^{\rm AB} + \zeta \, 
\theta^2 + \rho \, ({\mathbf J}_{\rm BH} - \sigma_{\rm BH} \, {\mathbf v})^2 \right] \, 
.
\ee
Here, $d/dt$ is a convective derivative (along the generators), $\tau \equiv 
\kappa^{-1}$ a characteristic time scale, $\eta$ the shear viscosity mentioned above, 
$\sigma_{\rm AB}$ the shear tensor of the 2-dimensional fluid motion, $\zeta = -\eta$ 
the bulk viscosity, $\theta$ the local rate of dilatation of the fluid motion, $\rho$ 
the (surface) resistivity, ${\mathbf J}_{\rm BH} - \sigma_{\rm BH} {\mathbf v}$ the 
horizon's conduction current (total current ${\mathbf J}_{\rm BH}$ minus convection 
current $\sigma_{\rm BH} {\mathbf v}$, where $\sigma_{\rm BH}$ is the horizon's 
surface charge density), and where $T_{\rm BH}$ can be interpreted as the {\it local} 
temperature of the horizon
\be
\label{eq1.20}
T_{\rm BH}  \equiv \frac{\kappa}{8 \pi \alpha} \, ,
\ee
where $\kappa$ (defined as above by $\ell^{\nu} \, \nabla_{\nu} \, \ell^{\mu} \equiv \kappa 
\, 
\ell^{\mu}$) is the local value of the surface gravity. [The surface gravity is 
uniform, on the horizon, for stationary black holes, but is generically non-uniform 
for evolving black holes.]

In the limit of an adiabatically slow evolution of the black-hole state, 
Eq.~(\ref{eq1.19}) reduces to the usual thermodynamical law giving the local increase 
of the entropy of a fluid element heated by the dissipations associated to viscosity 
and the Joule's law. [For non adiabatic evolutions, the L.H.S. of Eq.~(\ref{eq1.19}) 
can be interpreted, similarly to the famous a-causal Lorentz-Dirac 
radiation-reaction equation, in terms of an anticipated response of the black hole to 
external sollicitations \cite{Hartle,D79}.]

\section{Bekenstein's entropy and Hawking's temperature}

The striking similarity between the irreversible increase of the area of the horizon 
of a black hole, reviewed in the last Section, and the second law of thermodynamics,
motivated Bekenstein \cite{B72a,B72b,B73} to introduce the concept of black-hole 
entropy, say $S_{\rm BH}$. [See the Appendix  for
John Wheeler's account \cite{wheeler} of how Jacob Bekenstein started to think about
the concept of black-hole entropy and see Bekenstein's own account in \cite{bekensteinPT}.] 
He gave 
several arguments suggesting that $S_{\rm BH}$ is 
(as anticipated in the previous Section) proportional to the area $A$ of the black 
hole, and of the form
\be
\label{eq2.1}
S_{\rm BH} = \hat\alpha \, \frac{c^3}{\hbar G} \, A \equiv \hat\alpha \, 
\frac{A}{\ell_P^2} \, ,
\ee
where $\hat\alpha$ is a universal dimensionless number of order unity, and where 
$\ell_P \equiv \sqrt{\hbar G / c^3}$ denotes the Planck length. [We set Boltzmann's 
constant to unity.]

One of the arguments used by Bekenstein was an analysis of the {\it quantum 
limitations} on the existence of reversible transformations, in the sense of 
Christodoulou \cite{C70}. Indeed, from Eq.~(\ref{eq1.7n}) above 
the inequality (\ref{eq1.7}) can be saturated only 
in the limit where the particle captured by the black hole has {\it exactly} 
zero radial momentum
{\it on the horizon}. He considered that, in order of magnitude, quantum 
effects impose a minimal (proper) size, for a particle of mass $\mu$, of order its 
Compton wavelength $\lambda_c = \hbar / \mu c$, so that a particle can never be 
located exactly on the horizon, but only within an uncertainty $\sim \lambda_c$ in 
proper distance. Bekenstein then found that this uncertainty implied the existence of 
a lower bound on the difference between the L.H.S. and the R.H.S. of 
Eq.~(\ref{eq1.7}). When estimating the corresponding lower bound on the increase of 
the area of a black hole as it captures a particle, he found a value of order $\hbar G 
/ c^3 = \ell_P^2$, seemingly independent of all the characteristics of both the black 
hole and the particle.

Considering that, as a particle goes down a black hole, one looses (because of the 
no-hair theorems) an information of the order of one bit, and following Brillouin in 
identifying information with negative entropy \cite{B56}, he estimated that the 
dimensionless constant $\hat\alpha$ in Eq.~(\ref{eq2.1}) was probably numerically 
$\simeq (\ln 2) / (8\pi)$. In other words, Bekenstein's black-hole entropy was viewed 
by him as a measure of the information about the interior of a black hole which is 
inaccessible to an external observer. He gave several other arguments leading to a 
result of the form (\ref{eq2.1}), including an argument based on the analysis of 
quantum limitations on the efficiency of Carnot cycles using a black hole as heat 
sink. The latter argument suggested the need to attribute to a black hole a 
temperature of order of that corresponding (in the sense of 
Eqs.~(\ref{eq1.14})--(\ref{eq1.16}) above, with $\alpha \equiv \hat\alpha \, c^3 / 
(\hbar G)$) to the entropy (\ref{eq2.1}), {\it i.e.}
\be
\label{eq2.2}
T_{\rm BH} = \frac{1}{8 \pi \hat\alpha} \, \frac{\hbar}{c} \, \kappa \, ,
\ee
where $\kappa$ denotes the surface gravity of the black hole. [A factor $G/c^2$ 
disappeared between (\ref{eq2.1}) and (\ref{eq2.2}) because we measure here $\kappa$ 
in the usual way surface gravities are measured: $\kappa \sim GM / r^2 = c^2 / {\rm 
length}$.] Finally, Bekenstein conjectured the validity of a {\it generalized version 
of the second law of thermodynamics}, stating that the sum of the black-hole entropy 
(\ref{eq2.1}) and of the ordinary entropy in the exterior of the black hole never 
decreases.

Bekenstein's arguments could not yield a precise determination of the dimensionless 
coefficient $\hat\alpha$ in Eq.~(\ref{eq2.1}) or (\ref{eq2.2}) (nor could they really 
establish the universality of this coefficient). Soon after Bekenstein's original 
suggestions, Hawking (who, initially, did not believe that it made sense to attribute 
a non-zero temperature (\ref{eq2.2}) to a black hole) disco\-vered the universal 
phenomenon of quantum radiance of black holes \cite{H74,H75}. This work gave a precise 
meaning to a black-hole temperature of the form (\ref{eq2.2}). More precisely, Hawking 
found
\be
\label{eq2.3}
T_{\rm BH} = \frac{1}{2\pi} \, \frac{\hbar}{c} \, \kappa \, ,
\ee
which exactly corresponds to Eq.~(\ref{eq2.2}) if (and only if) the dimensionless 
coefficient $\hat\alpha$ there is equal to
\be
\label{eq2.4}
\hat\alpha = \frac{1}{4} \, .
\ee
In view of the importance of Hawking's result \cite{H74,H75}, let us (following 
\cite{DR76}) give a simplified derivation of the phenomenon of quantum radiance. Let 
us consider (for simplicity) a massless scalar field $\varphi (x)$ propagating in a 
Schwarzschild background. The quantum excitations of $\varphi (x)$ can be decomposed 
into classical mode functions, say
\be
\label{eq2.5}
\varphi_{\omega \ell m} (t,r,\theta,\varphi) = \frac{e^{-i\omega t}}{\sqrt{2 \pi \vert 
\omega \vert}} \, \frac{u_{\omega \ell m} (r)}{r} \, Y_{\ell m} (\theta , \varphi) \, 
,
\ee
which must satisfy
\be
\label{eq2.6}
0 = \Box_g \, \varphi = \frac{1}{\sqrt g} \, \partial_{\mu} (\sqrt g \, g^{\mu\nu} \, 
\partial_{\nu} \, \varphi) \, .
\ee

Before tackling the application of Eq.~(\ref{eq2.6}) to a Schwarzschild background, 
let us recall the essentials of the theory of particle creation by a background field. 
The quantum operator $\hat\varphi (x)$ describing (real, massless) scalar particles 
can be decomposed both with respect to some ``in'' basis of modes (describing free, 
incoming particles), say
\be
\label{eq2.n7}
\hat\varphi (x) = \sum_i \hat a_i^{\rm in} \, p_i^{\rm in} (x) + (\hat a_i^{\rm in})^+ 
\, n_i^{\rm in} (x) \, ,
\ee
and with respect to an ``out'' basis of modes (describing outgoing particles), say
\be
\label{eq2.n8}
\hat\varphi (x) = \sum_i \hat a_i^{\rm out} \, p_i^{\rm out} (x) + (\hat a_i^{\rm 
out})^+ \, n_i^{\rm out} (x) \, .
\ee
Here the $\hat a^{\rm in}$, $(\hat a^{\rm in})^+$, $\hat a^{\rm out}$, $(\hat a^{\rm 
out})^+$ are two sets of annihilation and creation operators (with, {\it e.g.}, $[\hat 
a_i^{\rm in} , (\hat a_j^{\rm in})^+] = \delta_{ij}$) corresponding to a decomposition 
in modes which can physically be considered as incoming positive-frequency ones 
$(p_i^{\rm in} (x))$, incoming negative-frequency ones ($n_i^{\rm in} (x)$; which can 
be taken to be the complex conjugate of $p_i^{\rm in} (x)$ in our case), outgoing 
positive-frequency ones $(p_i^{\rm out} (x))$ and outgoing negative-frequency ones 
$(n_i^{\rm out} (x))$.  These mode functions are normalized so that, say,
$(p_i^{\rm in} , p_j^{\rm in}) = \delta_{ij}$, $(p_i^{\rm in} , n_j^{\rm in}) = 0$,
$(n_i^{\rm in} , n_j^{\rm in}) = - \delta_{ij}$, where $(\varphi_1,\varphi_2)$ denotes
the standard (conserved) Klein-Gordon scalar product.

The ``in'' vacuum is defined by $\hat a_i^{\rm in} \vert {\rm 
in} \rangle = 0$, and the mean number of $i$-type ``out'' particles present in the 
in-vacuum is given by
\be
\label{eq2.n9}
\langle N_i \rangle = \langle {\rm in} \vert \, (a_i^{\rm out})^+ \, a_i^{\rm out} \, 
\vert {\rm in} \rangle = \sum_j \vert T_{ij} \vert^2
\ee
where $T_{ij} \equiv (p_i^{\rm out} , n_j^{\rm in})$ is the (standard Klein-Gordon) 
scalar product, {\it i.e.} physically the {\it transition amplitude}, between the 
incoming negative-frequency mode $n_j^{\rm in}$ and the outgoing positive-frequency 
one $p_i^{\rm out}$. In the black-hole case the in-vacuum will be defined by focussing 
on high-frequency wave packets propagating in the vicinity of the horizon ${\mathcal 
H}$, say some $n_j^{\rm in} (x)$, and frequency-analyzed in a local freely-falling 
frame. We will then be interested in the transition amplitude between such an incoming 
$n_j^{\rm in} (x)$ and an outgoing mode which reaches (future) infinity as a 
positive-frequency wave-packet (in the usual sense of the decomposition (\ref{eq2.5}) 
there). 

Inserting the explicit form (\ref{eq1.1}) of the Schwarzschild metric into 
Eq.~(\ref{eq2.6}), using the mode decomposition (\ref{eq2.5}), and introducing the 
usual ``tortoise'' radial coordinate \cite{MTW}
\be
\label{eq2.7}
r_* = \int \frac{dr}{1 - \frac{2M}{r}} = r + 2M \ln \left( \frac{r-2M}{2M} \right) \, 
,
\ee
leads to the following radial equation for $u_{\omega \ell m} (r)$:
\be
\label{eq2.8}
\frac{\partial^2 u}{\partial r_*^2} + (\omega^2 - V_{\ell} \, [r (r_*)]) \, u = 0 \, ,
\ee
where the effective radial potential $V_{\ell}$ reads
\be
\label{eq2.9}
V_{\ell} (r) = \left( 1 - \frac{2M}{r} \right) \left( \frac{\ell (\ell + 1)}{r^2} + 
\frac{2M}{r^3} \right) \, .
\ee
In the coordinate $r_*$, the horizon ($r=2M$) is located at $r_* \to -\infty$, {\it 
i.e.} in the left asymptotic region of the potential $V_{\ell} \, [r(r_*)]$, where 
$V_{\ell}$ tends exponentially towards zero. [Note that $V_{\ell} \, [r(r_*)]$ tends 
to zero both as $r_* \to -\infty$ ($r \to 2M$) and as $r_* \to +\infty$ ($r \to 
+\infty$), and creates a positive potential barrier around $r \sim 3M$.] Neglecting 
the potential $V_{\ell}$ in Eq.~(\ref{eq2.8}), we see that, near the horizon, the 
radial function $u(r_*)$ behaves as $e^{\pm i \omega r_*}$, so that (after factoring 
$1/r$ and $Y_{\ell m} (\theta,\varphi)$) the mode function (\ref{eq2.5}) behaves there 
as
\be
\label{eq2.10}
\varphi_{\omega} (t,r) \propto e^{-i \omega (t \pm r_*)} \, .
\ee

At this point we need to recall that the ``singularity'' of the Schwarzschild metric 
(\ref{eq1.1}) on the horizon $r=2M$ can be eliminated by introducing new coordinates, 
better adapted to the intrinsically regular geometry of the (future) horizon 
${\mathcal H}$. As above a simple choice are the (ingoing) Eddington-Finkelstein 
coordinates 
$(v,r)$ with the transformation $(t,r) \to (v,r)$ given by $v \equiv t + r_*$, $r 
\equiv r$ [see, {\it e.g.}, \cite{MTW}]. Using these coordinates, let us write down 
the expression, near ${\mathcal H}$ ($r=2M$, $v$ finite), of the mode (\ref{eq2.10}) 
describing a wave which is {\it outgoing} from ${\mathcal H}$, and which will have 
(when it reaches infinity) a {\it positive} frequency.

It is given by (\ref{eq2.10}) with a {\it minus} sign in front of $r_*$, and with 
$\omega > 0$. Using
\be
\label{eq2.11}
t - r_* = t + r_* - 2 r_* = v - 2 r_* = v - 2r - 4M \ln \left( \frac{r-2M}{2M} \right)
\ee
this outgoing mode $\varphi_{\omega}^{\rm out} \propto e^{-i \omega (t-r_*)}$ reads
\be
\label{eq2.12}
[ \varphi_{\omega}^{\rm out} (v,r)]_{\mathcal H}^{\rm near} \propto e^{-i \omega v} \, 
e^{+2 i \omega r} \left( \frac{r-2M}{2M} \right)^{i 4 M \omega} \, .
\ee
Because of the last factor on the R.H.S. of Eq.~(\ref{eq2.12}), this outgoing mode 
seems to be rather singular on ${\mathcal H}$: (i) it piles up an infinite number of 
oscillations of shorter and shorter wavelengths as $r \to 2M + 0$, and (ii) it is 
undefined inside the black hole, {\it i.e.} for $r < 2M$. Let us now {\it extend} the 
definition of the mode $\varphi_{\omega}^{\rm out}$ so that it defines, in a local 
(regular) Fermi-frame in the vicinity of ${\mathcal H}$, a high-frequency wave packet 
made only of (locally) {\it negative frequencies}. (This will define a mode $n_j^{\rm 
in} (x)$ in the sense of Eq.~(\ref{eq2.n7}).) Consider, in Minkowski space, a wave 
packet
\be
\label{eq2.13}
\varphi_- (x) = \int_{{\mathcal C}^-} d^4 k \, \tilde\varphi (k) \, 
e^{ik_{\mu} x^{\mu}}
\ee
made only of {\it negative frequencies}, {\it i.e.} such that the 4-momenta $k^{\mu}$ 
entering (\ref{eq2.13}) are all contained in the {\it past light cone} ${\mathcal 
C}^-$ of $k^{\mu}$. A convenient technical criterion for characterizing, in $x$-space, 
such a negative-frequency wave packet is the well-known condition that $\varphi_- (x)$ 
be analytically continuable to {\it complexified} spacetime points $x^{\mu} + i \, 
y^{\mu}$ with $y^{\mu}$ lying in the {\it future} light cone, $y^{\mu} \in {\mathcal 
C}^+$. This local, $x$-space criterion can be applied (in a local frame near 
${\mathcal H}$) to the wave packet (\ref{eq2.12}). As the infinitesimal displacement 
$r \to r - \varepsilon$, $v \to v$ is seen to be future directed (and null), this 
criterion finally tells us that the following new, extended wavepacket (defined by 
applying the analytic continuation $r \to r - i {\varepsilon}$ to (\ref{eq2.12}))
\be
\label{eq2.14}
n_{\omega} (v,r) \equiv N_{\omega} \, \varphi_{\omega}^{\rm out} (v,r - 
i {\varepsilon}) \propto e^{-i \omega v} \, e^{2i \omega r} \left( \frac{r - 2M - 
i {\varepsilon}}{2M} \right)^{i 4 M \omega}
\ee
is (when Fourier analyzed in the vicinity of ${\mathcal H}$) a {\it negative 
frequency} wave packet. Note that the new wavepacket $n_{\omega} (v,r)$ is defined on 
both sides of ${\mathcal H}$ and that we have included a new normalizing factor 
$N_{\omega}$ in its definition in terms of the analytic continuation of the ``old'' 
mode $\varphi_{\omega}^{\rm out}$ (which had its own normalization).

The normalization factor $N_{\omega}$ will give the main effect of the Hawking 
radiance and is simply determined as follows. The original modes (\ref{eq2.5}) (which 
vanished inside ${\mathcal H}$) were normalized so that the invariant scalar product 
defined by the (massless) Klein-Gordon equation took the form $(\varphi_{\omega_1 
\ell_1 m_1}, \varphi_{\omega_2 \ell_2 m_2}) = + \delta (\omega_1 - \omega_2) \, 
\delta_{\ell_1 \ell_2} \, \delta_{m_1 m_2}$. The analytic conti\-nuation $r \to r - 
i {\varepsilon}$ introduces, via the rotation by $e^{-i\pi}$ of $r-2M$ in 
$(r-2M)^{i4M\omega}$, a factor $(e^{-i\pi})^{i4M\omega} = e^{+4\pi M \omega}$ in the 
left part ($r < 2M$) of $n_{\omega}$. In other words
\be
\label{eq2.15}
n_{\omega} (r) = N_{\omega} [\theta (r-2M) \, \varphi_{\omega}^{\rm out} (r-2M) + 
e^{4\pi M \omega} \, \theta (2M-r) \, \varphi_{\omega}^{\rm out} (2M-r)] \, ,
\ee
where $\theta (x)$ is the step function. Computing the scalar product $ (
n_{\omega_1 \ell_1 m_1} , n_{\omega_2 \ell_2 m_2} )$ one gets, besides the same 
delta functions as above, a factor $\vert N_{\omega} \vert^2 \, [1 - (e^{4 \pi M 
\omega})^2]$, where the minus sign is due to the (essentially) negative-frequency 
aspect of $\varphi_{\omega}^{\rm out} (2M-r) \, \theta (2M-r)$.

The correct normalization (for a negative-frequency mode), $(n_{\omega_1 \ell_1 m_1} , 
n_{\omega_2 \ell_2 m_2}) = - \delta (\omega_1 - \omega_2) \, \delta_{\ell_1\ell_2} \, 
\delta_{m_1m_2}$ is then obtained for
\be
\label{eq2.16}
\vert N_{\omega} \vert^2 = \frac{1}{e^{8 \pi M \omega} - 1} \, .
\ee

\begin{figure}[ht]
\begin{center}
\includegraphics[scale=.8]{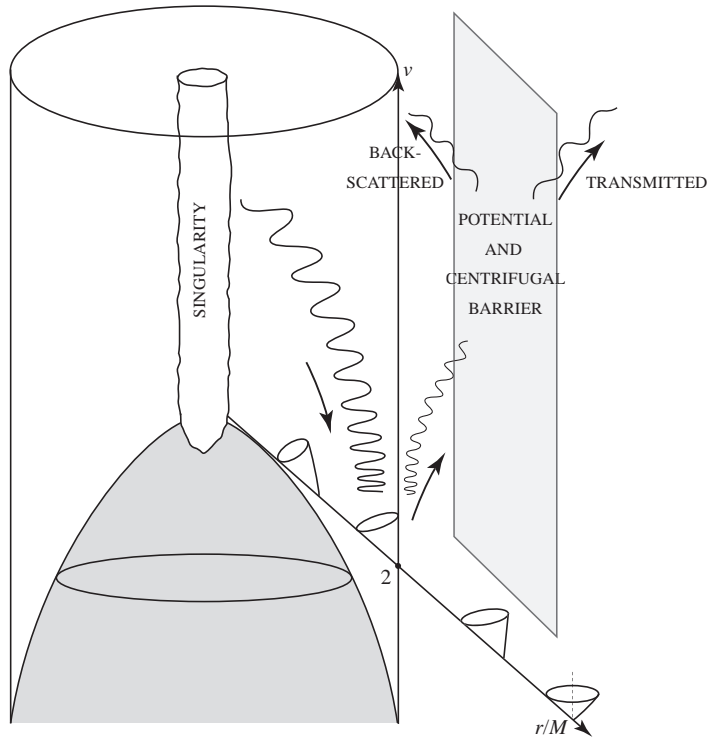}
\vglue 5mm
\caption{Splitting of an initial negative-frequency mode straddling the horizon
into a mode falling into the black hole, and an outgoing mode which,
after being partially reflected back into the black hole by the potential barrier
representing gravitational and centrifugal effects, ends up as  positive-frequency
Hawking radiation at infinity. The  antiparticle mode falling into
the black hole can be interpreted as a particle travelling backwards in time,
from the singularity down to the horizon [25] (hence the downwards
orientation of the arrow).}
\label{fig2}
\end{center}
\end{figure}

Finally, the result (\ref{eq2.15}) says that the initial negative-frequency mode 
$n_{\omega}$ straddling the horizon (and looking as a high-frequency wavepacket there) 
splits, as time evolves, into an outgoing mode $\varphi_{\omega}^{\rm out} (r-2M)$ 
(which will appear as having the positive-frequency $\omega$ at infinity), and another 
mode $\varphi_{\omega}^{\rm out} (2M-r)$ which falls towards the singularity. By the 
general formalism of particle production in curved spacetime recalled above, the 
number of outgoing particles in the mode $\omega$, $\ell$, $m$ is given by the sum 
(\ref{eq2.n9}) where $T_{ij}$ is the scalar product between $n_j^{\rm in} = 
n_{\omega_1 \ell_1 m_1}$ and $p_i^{\rm out} \propto \varphi_{\omega \ell m}^{\rm 
out}$. Actually, the mode $p_{\omega}^{\rm out}$, normalized at infinity, differs from 
the mode $\varphi_{\omega}^{\rm out}$ used above, and normalized on the horizon, by 
the effect of the potential barrier $V_{\ell} (r)$ in Eq.~(\ref{eq2.8}) (which was 
negligible near $r=2M$). If $V_{\ell} (r)$ were absent the looked-for transition 
amplitude would be 
\be
\label{eq2.17}
(\varphi_{\omega \ell m}^{\rm out} , n_{\omega_1 \ell_1 m_1}) = N_{\omega} \, \delta 
(\omega - \omega_1) \, \delta_{\ell \ell_1} \, \delta_{mm_1} \, ,
\ee
and the number of created particles (\ref{eq2.n9}) would contain $\vert N_{\omega} 
\vert^2$ times the square of $\delta (\omega - \omega_1)$, which, by Fermi's Golden 
Rule, is simply $\delta (\omega - \omega_1) \times \int dt / 2\pi$. Finally, adding 
the {\it grey body} factor $\Gamma_{\ell} (\omega)$, giving the fraction of the flux 
of $\varphi_{\omega}^{\rm out}$ which ends up at infinity because of the effect of 
$V_{\ell} (r)$ in Eq.~(\ref{eq2.8}) (see Fig.~\ref{fig2}), we get a {\it rate} of 
particle creation by the black hole of the form
\be
\label{eq2.18}
\frac{d \langle N \rangle}{dt} = \sum_{\ell , m} \int \frac{d\omega}{2\pi} \, \vert 
N_{\omega} \vert^2 \, \Gamma_{\ell} (\omega) = \sum_{\ell , m} \int 
\frac{d\omega}{2\pi} \, \frac{\Gamma_{\ell} (\omega)}{e^{8\pi M\omega} - 1} \, .
\ee

The formula (\ref{eq2.18}) exhibits, in the simple case of a Schwarzschild black hole, 
the Hawking phenomenon: a black hole radiates as if it were a black body of 
temperature $T_{\rm BH} = 1 / (8\pi M)$, covered up by a ``blanket'' with transmission 
factor $\Gamma_{\ell} (\omega)$. [The blanket being due to the effect of the potential 
$V_{\ell} (r)$ in Eq.~(\ref{eq2.8}), {\it i.e.} physically to the combined effect of 
gravitational attraction and centrifugal forces on the waves outgoing from the 
horizon.] For a Schwarzschild black hole the surface gravity is $\kappa = GM / r_S^2 = 
1 / (4M)$ (with $c = G = 1$), so that the Planck spectrum present in (\ref{eq2.18}) 
agrees with the temperature (\ref{eq2.3}). Note that the technical origin of a Planck 
spectrum with temperature $1/8\pi M$ is the logarithm $-4M\ln ((r-2M)/2M)$ in 
Eq.~(\ref{eq2.11}) which is itself linked to the logarithm $+2M \ln ((r-2M)/2M)$ in 
Eq.~(\ref{eq2.7}). In other words, it is the logarithmically divergent link between 
the coordinate time of a local freely falling frame near ${\mathcal H}$ and the time 
at infinity which is the source of a Planck spectrum.  Note that
the calculation above  technically relies on using such a logarithmically divergent
link between the negative-frequency wave-packet near the horizon,
and the positive-frequency wave-packet at infinity. When considering finite-frequency
packets at infinity, this formally means that one has worked with packets containing
{\it arbitrarily high} frequencies near the horizon.

When generalizing the analysis 
to a general black hole, one finds that the factor $4M$ gets replaced by the inverse 
of the surface gravity of the hole. One then finds a Planck-like spectrum containing 
(for bosons), besides a grey-body factor $\Gamma$, a factor 
$(e^{2\pi(\omega-p_{\varphi} \Omega - e\Phi) / \kappa} - 1)^{-1}$ exhibiting the 
combined effect of the general temperature (\ref{eq2.3}) and of the couplings of the 
conserved angular momentum $p_{\varphi}$ and electric charge $e$ to, respectively, the 
angular velocity $\Omega$ and electric potential $\Phi$ of the hole, see 
Eq.~(\ref{eq1.15}).

Numerically, $T_{\rm BH} \sim 10^{-6} \, K \, M_{\odot} / M$ so that Hawking's 
radiation is not astrophysically relevant for stellar-mass or larger black holes. 
Conceptually, it is, however, a beautiful discovery which connects relativistic 
gravity and quantum mechanics.

\section{Black holes, quantum mechanics and string theory}
\subsection{Conceptual puzzles}
The discovery of Hawking radiation led to new conceptual challenges. First, the 
obtention of a Planck-like spectrum with the temperature (\ref{eq2.3}) fixed the 
dimensionless coefficient $\hat\alpha$ in the general form (\ref{eq2.2}) to the value 
$\hat\alpha = \frac{1}{4}$. By the general ``first law of black-hole thermodynamics'' 
(\ref{eq1.15}) this would then correspond to an entropy (\ref{eq2.1}) with the same 
value of $\hat\alpha$, {\it i.e.}
\be
\label{eq3.1}
S_{\rm BH} = \frac{1}{4} \, \frac{c^3}{\hbar G} \, A \equiv \frac{1}{4} \, 
\frac{A}{\ell_P^2} \, .
\ee
However, Hawking's derivation of a temperature does not help in understanding the 
physical meaning of the Bekenstein-Hawking ``entropy'' (\ref{eq3.1}). Beyond the 
original hints of Bekenstein that $S_{\rm BH}$ is a measure of the information loss 
about what went into the black hole, it remains the challenge of interpreting in a 
precise way $S_{\rm BH}$, \`a la Boltzmann, as the logarithm of the number of quantum 
micro-states of a macroscopic black hole.

Another challenge comes from considering the end-point of the Hawking quantum 
eva\-poration process. In order of magnitude (and in units $c = \hbar = G = 1$), 
Hawking's radiation corresponds to an outgoing flux of energy of order (by Stephan's 
law) $\sim AT_{\rm BH}^4 \sim R_{\rm BH}^2 \, T_{\rm BH}^4 \sim M^2 \, M^{-4} \sim 
M^{-2}$. This loss should correspond to a secular decrease of the mass $M$ of the 
black hole so that $dM / dt \sim -M^{-2}$, {\it i.e.} $M^3 \sim M_0^3 - (t - t_0)$. 
This leads to a finite evaporation lifetime for a black hole of order
\be
\label{eq3.2}
t_{\rm evap} \sim t_P \left( \frac{M}{M_P} \right)^3 \sim 10^{-44} s \left( 
\frac{M}{10^{-5} g} \right)^3 \sim 10^{10} {\rm yr} \left( \frac{M}{10^{14} g} 
\right)^3
\ee
where $t_P \equiv (\hbar G / c^5)^{1/2} \simeq 5.4 \times 10^{-44} s$ is the Planck 
time and $M_P \equiv (\hbar c / G)^{1/2} \simeq 2.2 \times 10^{-5} g$ the Planck mass. 
Besides the (remote) astrophysical possibility \cite{H74} of seing today in the sky 
the explosions of black holes of initial mass $M \sim 10^{14} g$ formed in the big 
bang, the main consequence of the finiteness of the evaporation time (\ref{eq3.2}) is 
to oblige us to face the following questions: (i) what is the end point of the 
evaporation process?, and (ii) is there an ``information loss'' paradox due to the 
eventual final disappearance of the part of the wave packets in Fig.~\ref{fig2} 
which fell into the hole, and the concomitant emission of an outgoing flux of 
uncorrelated thermal radiation.  Hawking suggested \cite{H76} that if the quantum
radiance phenomenon ultimately leads to the disappearance of the black hole,
then the information about what has fallen in is completely lost, and one must
describe the evaporation process by a density matrix rather than by a unitary
quantum evolution.  In addition, we saw above that the calculation of the Hawking
process formally involved {\it arbitrarily high} frequencies near the horizon.
Many authors felt uncomfortable about this intermediate use of 
arbitrarily high (so-called ``trans-Planckian'') frequencies.
A lot of work was devoted to these issues, as well as 
to the issue of understanding the statistical meaning of the Bekenstein-Hawking
entropy.

We will not try here to summarize all the work done on these issues. Let us mention the 
existence of various lines of work, and give entries to the literature. Several authors 
explored the possibility that quantum black holes have a discrete mass spectrum, {\it 
e.g.} with the area $A = 16 \pi M_{\rm irr}^2$ having an uniformly spaced spectrum of 
eigenvalues $\frac{1}{4} \, A = \ell_P^2 (n + {\rm cst.})$. Then, to give a statistical 
meaning to the entropy $S_{\rm BH}$ one needs to assume that each discrete energy level 
$A_n$ has a huge degeneracy $g_n = \exp \left( \frac{1}{4} \, A_n / \ell_P^2 \right)$. See 
Refs.~\cite{B74,M86,K86,M94,B97}. Some authors have tried to identify the microstates 
responsible for black-hole entropy as near-horizon excitations \cite{ZT85,C95,S98,C99}. 
For entries in the literature on the ``black-hole information puzzle'', see 
\cite{H76,BOS93,P93,ALZP95,G95,LT99}. 
For discussions of the insensitivity of the Hawking process to the
arbitrarily high frequencies that seem to play a crucial role in all its derivations,
see \cite{FH90,U95,CT96,HB96}.
For suggestions that the so-called Loop Quantum Gravity may account 
for black-hole entropy see \cite{ABCK98,ABK00}. For a possible link between black-hole 
entropy and the quasinormal modes of black holes see \cite{Hod,Dreyer,Motl}.
Finally, let us mention the suggestion of a (black-hole motivated) {\it upper bound}, on 
the
entropy of any system, of the form $ S < \beta E R/ \hbar c$, where $\beta$ is
 a constant of order unity, $E$ the energy of the system, and $R$ its (effective) radius
 \cite{Bek81}.  See, however, \cite{MS03} and references therein,
 for a discussion of possible loopholes
 in this, and other (``holographic''-type) entropy bounds.
 
\subsection{Extremal black holes and supersymmetric solitonic states in string theory}
\subsubsection{BPS extremal black holes and D branes}
Here we shall focus on the most striking ``explanation'' of black-hole entropy, which was 
obtained within the framework of string theory. For early suggestions of a link between 
black-hole entropy and ``string entropy'' ({\it i.e.} the huge degeneracy of very massive 
string states) see \cite{BSW87,V86,S93}. After a seminal paper of Sen \cite{S95}, a 
breakthrough was brought by the works of Strominger and Vafa \cite{SV96} and Callan and 
Maldacena \cite{CM96}. These works on the microscopic origin of the Bekenstein-Hawking 
entropy of certain (Bogomolnyi-Prasad-Sommerfield; BPS) {\it extremal} black holes 
initiated a huge activity, which is  reviewed in \cite{Peet,DMW02}.  For textbook reviews 
see section 14.8
in volume 2 of \cite{Polbook} and  chapter 17 of \cite{Johnson}
 (see also \cite{Myers01} for an entry into the
literature on stringy black holes).

The line of work just mentioned on the micro-structure of certain BPS extremal black holes 
has led to very beautiful results. The factor $\frac{1}{4}$~\footnote{Though Section~II 
above discussed only 4-dimensional black holes, it has been shown that the factor 
$\hat\alpha = \frac{1}{4}$ is independent of the dimension. Note that many 
(but not all \cite{HLM96}) string 
calculations deal with higher-dimensional black holes, {\it e.g.} 5-dimensional ones, 
after compactification of 5 spatial dimensions.} in the entropy (\ref{eq3.1}) has been 
verified in the explicit counting of the degeneracy of some {\it extremal} black holes, and 
one could also verify some aspects of the physics of {\it near-extremal} holes, such as 
explaining Hawking radiation as a process of quantum decay of some unstable states 
\cite{CM96,DM96}, and even verifying the presence of the correct grey-body factor 
$\Gamma_{\ell} (\omega)$ in Eq.~(\ref{eq2.18}) (at least in the long-wavelength limit) 
\cite{MS97,DMW99}. Note that the Hawking temperature (\ref{eq2.3}), being proportional to 
the 
surface gravity $\kappa$, Eq.~(\ref{eq1.17}), {\it vanishes} in the limit of {\it extremal} 
black holes, {\it i.e.} black holes saturating the inequality (\ref{eq1.5}). [The BPS 
black holes automatically saturate this inequality.] One therefore expects any quantum 
representation of an extremal black hole to be a stable (non radiating) object. This is the 
case of the supersymmetric BPS states studied in the above references. Then, allowing for a 
small 
deviation from the BPS condition, and therefore a small deviation from extremality, is 
expected to lead to a weakly unstable object, which can be treated by perturbation theory. 
[Even so, the problem is more tricky than was first thought \cite{DMW02}.] Let us also 
mention that the study of extremal BPS black holes in certain limits led Maldacena 
\cite{M98} to conjecture a correspondence between string theory in anti-de-Sitter (AdS) 
spacetimes (times a compact manifold) and certain Conformal Field Theories (CFT). In turn, 
the study of this correspondence led to new insights into black-hole physics and to a
confirmation that the black-hole evaporation process should be describable as a unitary
process in quantum mechanics (see \cite{AGMOO} for a review).

However, in spite of its striking successes the work on the string-theory modelling of BPS 
black holes has some drawbacks and limitations. The main point is that this approach 
describes the quantum micro-structure of BPS black holes in terms of a (nearly 
non-interacting) gas of massive, charged string solitons, called ``Dirichlet branes'',
or simply $D$-branes \cite{Polchinski,Polbook}. For a recent  extensive review
of $D$-branes see \cite{Johnson}.  These solitons come in various dimensionalities:
a $D$-$p$ brane denotes a $D$-brane which is extended in $p$ spatial dimensions,
e.g. a $D$-0 brane denotes a Dirichlet point particle, while a
 $D$-1 brane denotes a Dirichlet {\it string}.  These solitons carry  ``Ramond-Ramond'' 
 charges, {\it i.e.} they create electric-like or magnetic-like $p'$-form fields
 called Ramond-Ramond fields. [$p'$-form fields are generalizations of  Maxwell fields:
 instead of a (one-index) vector potential $A_{\mu}$, now called a 1-form field, one 
considers
$p'$-index antisymmetric potentials $A_{\mu_1 \cdots \mu_{p'}}$.  In a $\cal D$-dimensional
spacetime, a $D$-$p$ brane naturally couples to an electric-like $(p+1)$-form field,
or to a magnetic-like $({\cal D} -p-3)$-form field.]

Roughly 
speaking the interaction between these solitons vanishes because of a cancellation between 
gravitational attraction and electrostatic repulsion. [Recall that the non-spinning limit 
of the extremality constraint (\ref{eq1.5}) yields $\vert Q \vert = G^{\frac{1}{2}} M$ 
which corresponds exactly to the pairwise cancellation between $- \, G M_1 M_2 / r^{d-2}$ 
and $+ \, Q_1 Q_2 / r^{d-2}$ in space dimension $d$ .] Because of the BPS condition (which 
requires the 
preservation of part of the original maximal supersymmetry), the classical cancellation 
just mentioned extends also to quantum cancellations between couplings involving bosonic 
fields and 
couplings involving fermionic ones. These quantum cancellations then entail the existence 
of certain {\it non-renormalization theorems}, stating that certain, specific quantities 
do not depend on the value of the string coupling constant, say $g$. In turn, these 
non-renormalization theorems are crucial to the whole enterprise because of the following 
fact, that I have not yet spelled out. All the string calculations of supposedly ``black 
hole'' properties (such as the degeneracy of micro-states or the decay rate of nearly-BPS 
states) actually refer to a system of $D$-brane solitons in the limit $g \to 0$ where 
these solitons admit a string-theory description. Roughly speaking, the mass of those 
solitons scale like $g^{-1}$, while the gravitational constant scales like $g^2$. 
Therefore the $g \to 0$ limit corresponds to a limit where the product $GM \propto g$ tends 
to zero, 
which means physically that the ``Schwarzschild radius'' of the considered assembly of 
(nearly coincident) $D$-branes is much smaller than the string length $\ell_s \sim 
\sqrt{\alpha'}$. This is the limit where the assembly of $D$-branes is physically a 
``point particle'', rather than a classically describable ``black hole'', having an 
horizon much larger than $\ell_s$. 

\subsubsection{The example of $D$-1--$D$-5 branes in type IIB string theory}
 One of the most studied system
is a configuration of $N_1$ $D$-1 branes and $N_5$ $D$-5 branes in type IIB string theory,
carrying some momentum along the direction of the $D$-1 branes (which are all parallel
to each other
and embedded within the $D$-5 branes). One compactifies, \`a la  Kaluza-Klein, the 5 
spatial
dimensions along which lie the $D$-5 and the $D$-1 branes. 
More precisely, the compactified manifold is $T^4 \times S^1$ where the 4-torus  $T^4$
has volume $(2 \pi)^4 V$ and the circle $S^1$ has length $ 2 \pi R$.
This has the effect of
quantizing the (internal) momentum moving along the branes, say $P=N \hbar/R$,
where $R$ is the compactified radius along the  $D$-1 direction. Finally, the configuration
appears as being ``point-like'' in the 4 remaining uncompactified spatial dimensions,
and this ``point'' carries three types of  charges (quantized in integers): $N_1, N_5$ and 
$N$.
The total mass-energy of the configuration reads (in string units $\alpha' =1$)

\be
\label{Dmass}
M = \frac{N_1 R}{g} + \frac{N_5 R V}{g} + \frac{N}{R} \, .
\ee
$D$-brane techniques (together with the general Cardy formula \cite{Cardy}
giving the density of states in two-dimensional conformal field theory)
allow one to evaluate the quantum degeneracy $\cal D$ of the
(supersymmetric) ground state of this system of $D$-1 and $D$-5 branes  
\cite{SV96,CM96,DMW02}.
Indeed, though the full structure of those solitons cannot be described by
perturbative string theory, one {\it can} describe, in the limit $g \to 0$ where
the $D$-branes become infinitely massive, the excitations of collections
of $D$-branes in terms of (perturbative) {\it open strings} whose end points
are constrained to move on the (nearly fixed) $D$-brane submanifolds 
\cite{Polchinski,Polbook,Johnson}.
In the considered case, one finds that the excitations of the $D$-1, $D$-5 brane 
configuration
are described by a two-dimensional superconformal field theory with 
$4 N_1 N_5$ real scalars and, by supersymmetry, $4 N_1 N_5$ Majorana
fermions. This yields a central charge $c= 6 N_1 N_5$ for a system of 
left-movers\footnote{Only left movers are excited in sypersymmetric states, corresponding
to extremal BPS black holes.  By adding a little bit of excitation energy in
right movers one can then describe near supersymmetric $D$-brane configurations
corresponding to near extremal black holes \cite{CM96}.}
 with $L_0$ level given by $N$. The general Cardy formula
$ {\cal D}(N)   \sim  \exp (2 \pi \sqrt{c N/6})$
(asymptotically valid for large values of $N$; see \cite{C00} for power-law
corrections to the general exponential Cardy formula) then gives the following result
for the quantum degeneracy of the $D$-brane configuration
\be
\label{Dbrane}
{\cal D}  \sim  e^{2 \pi \sqrt{N_1 N_5 N}} \, ,
\ee
which will be compared below to the degeneracy of the corresponding black hole expected 
from the 
Bekenstein-Hawking entropy.

\begin{figure}[ht]
\begin{center}
\includegraphics[scale=.8]{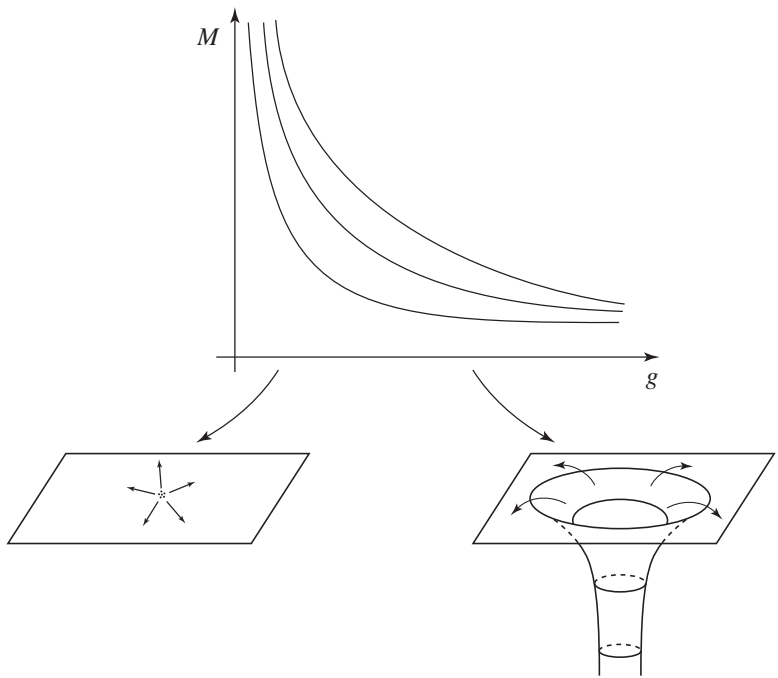}
\vglue 5mm
\caption{Evolution of the mass-energy, and of the physical structure, of an
initially point-like configuration of $D$-branes as the string coupling $g$
increases.}
\label{fig3}
\end{center}
\end{figure}

At some larger value of $g$ one expects the collection 
of $D$-branes to smoothly turn into a quasi-classical black hole.  However, the present 
state of development of string theory does not allow one to explicitly study this 
transition, and to set up a well-defined quantum Hilbert-space description of 
quasi-classical black holes. One has to crucially rely on non-renormalization theorems to 
``blindly'' continue the result of calculations well-defined as $g \to 0$ into physical 
properties of quasi-classical black holes (such as the degeneracy of a certain quantum BPS 
state which is explicitly describable when $g \to 0$, but whose ``physical form'' for 
larger $g$ cannot be concretely described as a quantum state).
For instance,  the $D$-1 $D$-5 system mentioned above is expected to appear,
when $g$ (or rather some effective value of $g$, say $g_{\rm eff} $,
which is a combination of $g$ and of the integers $N_1,N_5,N$) gets larger than
one, as a (BPS) black hole, {\it i.e.} a generalization of the extremal Reissner-Nordstr\o 
m 
black hole ($ G^{1/2} M = \vert Q \vert$), carrying the (conserved) quantized charges 
$N_1,N_5,N$.
In other words,  when  $g_{\rm eff}$ gets large one can no longer describe this
system of massive solitons as a ``point charge'', because the combined gravitational
effect of the total mass of the soliton configuration, and of the mass-energy associated to 
the
various Ramond-Ramond  (and internal-momentum related) fields it creates,
generates a strong curvature of space around it, and ultimately gives birth to a
finite area horizon, instead of a ``point with nothing inside it''.
For instance, the $(4+1)$-dimensional (Einstein-frame) metric generated, in the
uncompactified 
dimensions, by a $D$-1 $D$-5 system carrying the quantized charges $N_1$, $N_5$, $N$, reads
\be
\label{4n1}
ds_E^2 = - (f_1 \, f_5 \, f)^{-\frac{2}{3}} \, dt^2 + (f_1 \, f_5 \, f)^{\frac{1}{3}} \, 
(dx_1^2 + 
\ldots + dx_4^2) \, ,
\ee
with
\be
\label{4n2}
f_1 = 1 + \frac{4GR}{g\pi \alpha'} \, \frac{N_1}{r^2} \, , \ f_5 = 1 + \alpha' g \, 
\frac{N_5}{r^2} 
\, , \ f = 1 + \frac{4G}{\pi R} \, \frac{N}{r^2} \, ,
\ee
where $r^2 \equiv x_1^2 + \ldots + x_4^2$ denotes a (squared) radial coordinate in the 4 
uncompactified spatial dimensions. See  references
\cite{CM96,Peet,DMW02,Polbook,Johnson}
 for the explicit form of the other fields 
generated by a system of coincident $D$-branes.  We shall content ourselves
here with illustrating the evolution, with $g$, of the physical structure of
a  system of coincident $D$-branes  in Fig.~\ref{fig3}.

 The resulting extremal black hole is uniquely characterized
by the knowledge of the conserved quantized charges $N,N_1,N_5$. 
In particular, the horizon of the black-hole metric (\ref{4n1}) is located at the radial 
coordinate 
$r=0$. [Note that, as $r \to 0$, the time-time component of the metric (\ref{4n1}) is 
proportional 
to $((r^{-2})^3)^{-\frac{2}{3}} = r^{+4}$.] The ``area'' (which is that of a 3-dimensional 
sphere) 
of the black-hole horizon is finite, because $(f_1 \, f_5 \, f)^{\frac{1}{3}} \propto 
r^{-2}$ 
compensates the factor $r^2$ in the non-radial part of the spatial metric $r^2 \, d \, 
\Omega_3^2$. 
It is an easy matter to evaluate this ``area'', and thereby, using formula Eq.(\ref{eq3.1}) 
above, 
the corresponding Bekenstein-Hawking entropy. One finds that the dependences on $G$, $R$, 
$g$ and $\alpha'$ cancell out so that the latter entropy depends
only on the quantized charges carried by the BPS black hole, and equals

\be
S_{BH}= 2 \pi \sqrt{N_1 N_5 N} \, ,
\ee

which {\em exactly coincides with the  logarithm of the $D$-brane degeneracy}
 Eq.(\ref{Dbrane}).

\subsection{Schwarzschild black holes and  string states}
\subsubsection{String--black-hole transition and  conjectured correspondence between string  
and black-hole states}
Because of the drawback of not being able to follow in detail the ``transformation''
between a point-like system of $D$-branes and a black hole, we shall focus  henceforth
on another type of string-theory calculation of 
the properties of massive string states and on their transition to black-hole states 
\cite{HP97,HP98,DV00}. As just said, most of the stringy literature has concentrated on 
some special, supersymmetric extremal black holes (BPS black holes). These black holes  
carry special (Ramond-Ramond) charges and their microscopic structure seem to be 
describable (when  $g^2 < g_c^2$) in terms of Dirichlet-branes. By contrast, we consider 
here the simplest, Schwarzschild black holes (in any space dimension $d$). It will be 
argued that their ``microscopic structure'' at low $g^2$ involves only {\it fundamental 
string 
states} (instead of solitonic string states, such as $D$-branes).
 However, the lack of supersymmetry means that it becomes essential 
to deal with {\it self-gravity effects}.

To start with, let us recall that thirty years ago the study of the spectrum of string 
theory revealed~\cite{FVBM} a huge degeneracy of states growing as an exponential of the 
mass. This string result was obtained a few years before Bekenstein~\cite{B73} 
proposed that the entropy of a black hole should be proportional to the area of its 
horizon in Planck units, and Hawking~\cite{H75} fixed the constant of proportionality 
after discovering that black holes do emit thermal radiation at a temperature $T_{\rm Haw} 
\sim R_{\rm BH}^{-1}$.

When string and black-hole entropies are compared one immediately notices a striking 
difference: string entropy\footnote{As we shall discuss, the self-interaction of a string 
lifts the huge degeneracy of free string states. One then defines the entropy of a narrow 
band of string states, defined with some energy resolution $M_s \lesssim \Delta \, E \ll 
M$, as the logarithm of the number of states within the band $\Delta \, E$.} is 
proportional to the first power of mass in any number of spatial dimensions $d$, while 
black-hole entropy is proportional to a $d$-dependent power of the mass, always larger 
than $1$. In formulae:
\begin{equation}
S_s \sim {\alpha' M \over \ell_s} \sim M / M_s ~~~~ , ~~~~~~
S_{\rm BH} \sim \frac{{\rm Area}}{G} \sim \frac{R_{\rm BH}^{d-1}}{G} \sim \frac{(g^2 \, 
M / M_s)^{\frac{d-1}{d-2}}}{g^2}\; ,
\label{entropies}
\end{equation}
where, as usual, $\alpha'$ is the inverse of the classical string tension, $\ell_s \sim 
\sqrt{\alpha' \hbar}$ is the quantum length associated with it\footnote{Below, we shall 
use the precise definition $\ell_s \equiv \sqrt{2 \alpha' \hbar}$, but, in this  paragraph, 
we neglect factors of order unity.}, $M_s \sim \sqrt{\hbar / \alpha'}$ is the 
corresponding string mass scale, $R_{\rm BH}$ is the Schwarzschild radius associated with 
$M$:
\begin{equation}
R_{\rm BH} \sim (G \, M)^{1/(d-2)} \; ,\label{eq1.1bis}
\end{equation}
and we have used that, at least at sufficiently small coupling, the Newton constant and 
$\alpha'$ are related via the string coupling by
\begin{equation} 
G \sim g^2 (\alpha')^{(d-1)/2}
\label{eqGalpha}
\end{equation} 
(or more 
geometrically, $\ell_P^{d-1}\sim g^2 \ell_s^{d-1}$).

Given their different mass dependence, it is obvious that, for a given set of the  
fundamental constants $G, \alpha', g^2$, $S_s > S_{\rm BH}$ at sufficiently small $M$, 
while the opposite is true at sufficiently large $M$. Obviously, there has to be a 
critical value of $M$, $M_c$, at which $S_s = S_{\rm BH}$. This observation had led Bowick 
et al.~\cite{BSW87} to conjecture that large black holes end up their Hawking-evaporation 
process when $M = M_c$, and then transform into a higher-entropy string state without ever 
reaching the singular zero-mass limit. This reasoning is confirmed \cite{V86} by the 
observation that, in string theory, the fundamental string length $\ell_s$ should set a 
minimal value for the Schwarzschild radius of any black hole (and thus a maximal value for 
its Hawking temperature). It was also noticed~\cite{BSW87,S93,GVDivonne} that, precisely 
at $M= M_c$, $R_{\rm BH} = \ell_s$ and the Hawking temperature equals the Hagedorn 
temperature of string theory. For any $d$, the value of $M_c$ is given  by:
\begin{equation}
M_c \, \sim M_s g^{-2} \, . \label{eq1.2bis}
\end{equation}

Susskind and collaborators~\cite{S93,halyo} went a step further and proposed that the 
spectrum of black holes and the spectrum of single string states be ``identical'', in the 
sense that there be a one to one {\it correspondence} between (uncharged) fundamental 
string 
states and (uncharged) black-hole states. Such a ``correspondence principle'' has been 
generalized by Horowitz and Polchinski~\cite{HP97} to a wide range of charged black-hole 
states (in any dimension). Instead of keeping fixed the fundamental 
constants and letting $M$ evolve by evaporation, as considered above, one can 
(equivalently) describe the physics of this conjectured correspondence by following a 
narrow band of states, on both sides of and through, the string $\rightleftharpoons$ black 
hole transition, by keeping fixed the entropy\footnote{One uses here the fact that, during 
an adiabatic variation of $g$, the entropy of the black hole $S_{\rm BH} \sim ({\rm Area}) 
/ G \sim R_{\rm BH}^{d-1} / G$ stays constant. This result (known to hold in the 
Einstein conformal frame) applies also in string units because 
$S_{\rm BH}$ is dimensionless.} $S = S_s = S_{\rm BH}$, while  adiabatically\footnote{The 
variation of $g$ can be seen, depending on one's taste, either as a real, adiabatic change 
of $g$ due to a varying dilaton background, or as a mathematical way of following energy 
states.} varying the string coupling $g$, {\it i.e.} the ratio between $\ell_P$ and 
$\ell_s$. The correspondence principle then means that if one increases $g$ each (quantum) 
string state should turn into a (quantum) black-hole state at sufficiently strong 
coupling, while, conversely, if $g$ is decreased, each black-hole state should 
``decollapse'' and transform into a string state at sufficiently weak coupling. For all 
the reasons mentioned above, it is very natural to expect that, when starting from a black 
hole state, the critical value of $g$ at which a black hole should turn into a string is 
given, in clear relation to
(\ref{eq1.2bis}), by
\begin{equation}
g_c^2 \, M \sim M_s \, , \label{eq1.2'}
\end{equation}
and is related to the common value of string and black-hole entropy 
via
\begin{equation}
g_c^2 \sim \frac{1}{S_{\rm BH}} = \frac{1}{S_s}\; . \label{eq1.2''}
\end{equation}
Note that $g_c^2 \ll 1$ for the very massive states ($M \gg M_s$) that we consider. This 
justifies our use of the perturbative relation (\ref{eqGalpha}) between $G$ and $\alpha'$.

As we said above, in the case of extremal BPS, and nearly extremal, black holes
the conjectured correspondence was dramatically confirmed through the work of Strominger 
and Vafa~\cite{SV96} and others \cite{S95,CM96,DMW02} leading to a statistical mechanics 
interpretation of black-hole entropy in terms of the number of microscopic states sharing 
the same macroscopic quantum numbers. However, little is 
known about whether and how the correspondence works for non-extremal, non BPS black 
holes, such as the simplest Schwarzschild black hole\footnote{For simplicity, we shall
consider in this work only Schwarzschild black holes, in any number $d \equiv D-1$ of 
non-compact spatial dimensions.}. By contrast to BPS states whose mass is protected by 
supersymmetry, we shall consider here the effect of varying $g$ on the mass and size of 
non-BPS string states. 

Although it is remarkable that black-hole and string entropy coincide when $R_{\rm BH} = 
\ell_s$, this is still not quite sufficient to claim that, when starting from a string 
state, a string becomes a black hole at $g = g_c$. In fact, the process in which one 
starts from a string state in flat space and increases $g$ poses a serious 
puzzle~\cite{S93}. Indeed, the radius of a typical excited string state of mass $M$ is 
generally thought of being of order
\begin{equation}
R_s^{\rm rw} \sim \ell_s (M / M_s)^{1/2} \, , 
\label{eq1.5bis}
\end{equation}
as if a highly excited string state were a random walk made of $M/M_s = \alpha'M/\ell_s$ 
segments of length $\ell_s$~\cite{rw}. [The number of steps in this random walk is, as is 
natural, the string entropy (\ref{entropies}).] The ``random walk'' radius 
(\ref{eq1.5bis}) is much larger than the Schwarzschild radius for all 
couplings $g \le g_c$, or, equivalently, the ratio of self-gravitational binding energy to 
mass (in $d$ spatial dimensions)
\begin{equation}
\frac{G \, M}{(R_s^{\rm rw})^{d-2}} \sim \left( \frac{R_{\rm BH} (M)}{R_s^{\rm rw}} 
\right)^{d-2} \sim g^2 \left( \frac{M}{M_s} \right)^{\frac{4-d}{2}} 
\label{eq1.6bis}
\end{equation}
remains much smaller than one (when $d>2$, to which we restrict ourselves) up to, and 
including, the transition point. In view of (\ref{eq1.6bis}) it does not seem natural to 
expect that a string state will ``collapse'' to a black hole when $g$ reaches the value 
(\ref{eq1.2'}). One would expect a string state of mass $M$ to turn into a black hole only 
when its typical size is of order of $R_{\rm BH} (M)$ (which is of order $\ell_s$ at the 
expected transition point (\ref{eq1.2'})). According to Eq.~(\ref{eq1.6bis}), this seems 
to happen for a value of $g$ much larger than $g_c$.

Horowitz and Polchinski~\cite{HP98} have addressed this puzzle by means of a ``thermal 
scalar'' formalism~\cite{chi}. Their results suggest a resolution of the puzzle when $d=3$ 
(four-dimensional spacetime), but lead to a rather complicated behaviour when $d \geq 4$. 
Moreover, even in the simple $d=3$ case, the formal nature of the auxiliary ``thermal 
scalar'' renders unclear (at least to me) the physical interpretation of their analysis. 
In the next section, I will review the results of Ref.~\cite{DV00} whose aim was to clarify 
the 
string 
$\rightleftharpoons$ black-hole transition by a direct study, in 
real spacetime, of the size and mass of a {\it typical} excited string, within the 
microcanonical ensemble of {\it self-gravitating} strings. 

\subsubsection{A microcanonical approach to self-gravitating strings}
The results of~\cite{DV00} lead 
to a rather simple picture of the transition, in any dimension. 
Let us summarize them before entering into the technical details of the analysis.

The critical value for the transition is (\ref{eq1.2'}), 
or (\ref{eq1.2''}) in terms of the entropy $S$, for both directions of the string 
$\rightleftharpoons$ black hole transition. In {\it three spatial 
dimensions}, one finds that the size (computed in real spacetime) of a {\it typical 
self-gravitating} string is given by the random walk value (\ref{eq1.5bis}) when $g^2 \le 
g_0^2$, with $g_0^2 \sim (M/M_s)^{-3/2} \sim S^{-3/2}$, and by
\begin{equation}
R_{\rm typ} \sim \frac{1}{g^2 \, M} \, , 
\label{eq1.8bis}
\end{equation}
when $g_0^2 \le g^2 \le g_c^2$. Note that $R_{\rm typ}$ smoothly interpolates between 
$R_s^{\rm rw}$ and $\ell_s$. This result confirms the picture proposed by Ref.~\cite{HP98} 
when $d=3$, but with the bonus that Eq.~(\ref{eq1.8bis}) refers to a radius which is 
estimated directly in physical space (see below), and which is the size of a typical 
member of the microcanonical ensemble of self-gravitating strings. In all higher 
dimensions\footnote{With the proviso that the consistency of the analysis is open to doubt 
when $d\geq 8$.}, we find that the size of a typical self-gravitating string remains fixed 
at the random walk value (\ref{eq1.5bis}) when $g \le g_c$. However,  when $g$ gets close 
to a value of order $g_c$, the ensemble of self-gravitating strings becomes (smoothly in 
$d=4$, but suddenly in $d \geq 5$) dominated by very compact strings of size $\sim \ell_s$ 
(which are then expected to collapse with a slight further increase of $g$ because the 
dominant size is only slightly larger than the Schwarzschild radius at $g_c$). The results 
of~\cite{DV00} confirm and clarify the main idea of a correspondence between string states 
and black hole states~\cite{S93,halyo,HP97,HP98}, and suggest that the transition between 
these states is rather smooth, symmetrical with no apparent ``hysteresis effect'', 
and with continuity in 
entropy, mass, typical size, and luminosity. It is, however, beyond the technical grasp of 
this analysis to compute any precise number at the transition (such as the famous 
factor $1/4$ in the Bekenstein-Hawking entropy formula). Let us now enter into
some of the technical details.

For simplicity, we deal with open bosonic strings ($\ell_s \equiv \sqrt{2 \, \alpha'}$, 
$0 \leq \sigma \leq \pi$)
\begin{equation}
X^{\mu} (\tau , \sigma) = X_{\rm cm}^{\mu} (\tau , \sigma) + \widetilde{X}^{\mu} (\tau , 
\sigma) \, , 
\label{eq2.4bis}
\end{equation}
\begin{equation}
X_{\rm cm}^{\mu} (\tau , \sigma) = x^{\mu} + 2 \, \alpha' \, p^{\mu} \, \tau \, , 
\label{eq2.5bis}
\end{equation}
\begin{equation}
\widetilde{X}^{\mu} (\tau , \sigma) = i \, \ell_s \sum_{n \not= 0} \ 
\frac{\alpha_n^{\mu}}{n} \ e^{-i n \tau} \, \cos \, n \, \sigma \, .
\label{eq2.6bis}
\end{equation}
Here, we have explicitly separated the center of mass motion $X_{\rm cm}^{\mu}$ (with 
$[x^{\mu} , p^{\nu}] = i \, \eta^{\mu \nu}$) from the oscillatory one 
$\widetilde{X}^{\mu}$, which is expressed in terms of two parameters: the worldsheet 
time parameter $\tau$  and the worldsheet spatial parameter $\sigma$ ($[\alpha_m^{\mu} , 
\alpha_n^{\nu}] = m \, \delta_{m+n}^0 \, 
\eta^{\mu \nu}$). The free spectrum is given by $\alpha' \, M^2 = N-1$ where $(\alpha \cdot 
\beta\equiv \eta_{\mu \nu} \, \alpha^{\mu} \, \beta^{\nu} \equiv - \alpha^0\, \beta^0 + 
\alpha^i \, \beta^i)$
\begin{equation}
N = \sum_{n=1}^{\infty} \ \alpha_{-n} \cdot \alpha_n = \sum_{n=1}^{\infty} \ n \, N_n \, . 
\label{eq2.7bis}
\end{equation}
Here $N_n \equiv a_n^{\dagger} \cdot a_n$ is the occupation number of the $n^{\rm th}$ 
oscillator ($\alpha_n^{\mu} = \sqrt{n} \ a_n^{\mu}$,$[a_n^{\mu} , a_m^{\nu \dagger}] = 
\eta^{\mu \nu} \, \delta_{nm}$, with $n,m$ positive).

The decomposition (\ref{eq2.4bis})--(\ref{eq2.6bis}) holds in any conformal gauge 
($(\partial_{\tau} \, X^{\mu} \pm \partial_{\sigma} \, X^{\mu})^2 = 0$). One can further 
specify the choice of worldsheet coordinates by imposing
\begin{equation}
n_{\mu} \, X^{\mu} (\tau , \sigma) = 2 \alpha' (n_{\mu} \, p^{\mu}) \, \tau \, , 
\label{eq2.8bis}
\end{equation}
where $n^{\mu}$ is an arbitrary timelike or null vector ($n \cdot n \leq 
0$)~\cite{scherk}. Eq.~(\ref{eq2.8bis}) means that the $n$-projected oscillators $n_{\mu} 
\, \alpha_m^{\mu}$ are set equal to zero. As we shall be interested in quasi-classical, 
very massive string states ($N \gg 1$) it should be possible to work in the ``center of 
mass'' gauge, where the vector $n^{\mu}$ used in Eq.~(\ref{eq2.8bis}) to define the 
$\tau$-slices of the world-sheet is taken to be the 
total momentum $p^{\mu}$ of the string. This gauge is the most intrinsic way to describe a 
string in the classical limit. Using this intrinsic gauge, one can covariantly {\it define 
the proper rms size of a massive string state} as
\begin{equation}
R^2 \equiv \frac{1}{d} \ \langle (\widetilde{X}_{\perp}^{\mu} \, (\tau, \sigma))^2 
\rangle_{\sigma , \tau} \, ,
\label{eq2.9bis}
\end{equation}
where $\widetilde{X}_{\perp}^{\mu} \equiv \widetilde{X}^{\mu} - p^{\mu} (p \cdot 
\widetilde{X}) / (p \cdot p)$ denotes the projection of $\widetilde{X}^{\mu} \equiv 
X^{\mu} - X_{\rm cm}^{\mu} (\tau)$ orthogonally to $p^{\mu}$, and where the angular 
brackets denote the (simple) average with respect to $\sigma$ and $\tau$. 

In the center of mass gauge, $p_{\mu} \, \widetilde{X}^{\mu}$ vanishes by definition, and 
Eq.~(\ref{eq2.9bis}) yields simply
\begin{equation}
R^2 = \frac{1}{d} \, \ell_s^2 \, {\cal R} \, ,
 \label{eq2.10bis}
\end{equation}
with (after discarding a logarithmically infinite, but state independent, contribution)
\begin{equation}
{\cal R} \equiv \sum_{n=1}^{\infty} \ \frac{\alpha_{-n} \cdot \alpha_n}{n^2} =  
\sum_{n=1}^{\infty} \ \frac{a_n^{\dagger} \cdot a_n}{n}= \sum_{n=1}^{\infty} \ 
\frac{N_n}{n} \, . 
\label{eq2.11bis}
\end{equation}

We wish to estimate the {\it distribution function in size of the ensemble of free string 
states} of mass $M$, {\it i.e.} to count the number of string states, having some fixed 
values of $M$ and $R$ (or, equivalently, $N$ and ${\cal R}$). An approximate estimate of 
this number (``degeneracy'') is~\cite{DV00}
\begin{equation}
{\cal D} \, (M,R) \sim \exp \, [ c \, (R) \, a_0 \, M ] \, ,
\label{eq2.30}
\end{equation}
where $a_0 = 2 \, \pi \, ((d-1) \, \alpha' / 6)^{1/2}$ and
\begin{equation}
c \, (R) = \left( 1 - \frac{c_1}{R^2} \right) \left( 1 - c_2 \, \frac{R^2}{M^2} \right) \, 
, 
\label{eq2.31}
\end{equation}
with the coefficients $c_1$ and $c_2$ being of order unity in string units. The 
coefficient $c \, (R)$ gives the fractional reduction in entropy brought by imposing a 
size constraint. Note that (as expected) this reduction is minimized when $c_1 \, R^{-2} 
\sim c_2 \, R^2 / M^2$, {\it i.e.} for $R \sim R_{\rm rw} \sim \ell_s \, \sqrt{M/M_s}$.

We also need to estimate the {\it mass shift of string states} (of mass $M$ and size $R$) 
due to the exchange of the various long-range fields which are universally coupled to the 
string: graviton, dilaton and axion. As we are interested in very massive 
string states, $M \gg M_s$, in extended configurations, $R \gg \ell_s$, we expect that 
massless exchange dominates the (state-dependent contribution to the) mass shift. [The 
exchange of spin 1 fields (for open strings) becomes negligible when $M \gg M_s$ because 
it does not increase with $M$.]

The evaluation, in string theory, of (one loop) mass shifts for massive states is 
technically quite involved, and can only be tackled for the states which are near the 
leading Regge trajectory~\cite{mshift}. [Indeed, the vertex operators creating these 
states are the only ones to admit a manageable explicit oscillator representation.] As we 
consider states which are very far from the leading Regge trajectory, there is no hope of 
computing exactly (at one loop) their mass shifts. 

In Ref.~\cite{DV00} we could estimate the one-loop mass-shift by resorting to a 
semi-classical approximation. The starting point of this semi-classical approximation is 
the effective action of self-gravitating fundamental strings derived in Ref.~\cite{BD98}. 
Using coherent-state methods \cite{AABO,scherk,GSW} and a generalization of Bloch's 
theorem (see Eq.~(3.13) of~\cite{DV00}) one finds
\begin{equation}
\delta \, M \simeq - c_d \, G \, \frac{M^2}{R^{d-2}}\, , 
\label{eq3.25}
\end{equation}
with the (positive) numerical constant
\begin{equation}
c_d = \left[ \frac{d-2}{2} \, (4\pi)^{\frac{d-2}{2}} \right]^{-1} \, 
,\label{eq3.26}
\end{equation}
equal to $1 / \sqrt{\pi}$ in $d=3$.

The result (\ref{eq3.25}) was expected in order of magnitude, but it is important to check 
that it approximately comes out of a detailed calculation of the mass shift which 
incorporates both relativistic and quantum effects and which uses the precise definition 
(\ref{eq2.11bis}) of the squared size.

Finally, let us mention that, by using the same tools Ref.~\cite{DV00} has computed the 
imaginary part of the mass shift $\delta \, M = \delta \,M_{\rm real} - i \, \Gamma / 2$, 
{\it i.e.} the total decay rate $\Gamma$ in massless quanta, as well as the total power 
radiated $P$. In order of magnitude these quantities are
\begin{equation}
\Gamma \sim g^2 \, M \ , \ P \sim g^2 \, M \, M_s \, . 
\label{eq3.29}
\end{equation}

Finally one combines the results above, Eqs.~(\ref{eq2.30}) and 
(\ref{eq3.25}), and heuristically extend them at the limit of their domain of validity. 
We consider a narrow band of string states that we follow when increasing adiabatically 
the string coupling $g$, starting from $g = 0$. Let $M_0$, $R_0$ denote the ``bare'' 
values ({\it i.e.} for $g \rightarrow 0$) of the mass and size of this band of states. 
Under the adiabatic variation of $g$, the mass and size, $M$, $R$, of this band of 
states will vary. However, the entropy $S(M,R)$ remains constant under this adiabatic 
process: $S(M,R) = S (M_0 , R_0)$. We consider states with sizes $\ell_s \ll R_0 \ll M_0$ 
for which the correction factor,
\begin{equation}
c \, (R_0) \simeq (1 - c_1 \, R_0^{-2}) \, (1 - c_2 \, R_0^2 / M_0^2) \, ,
\label{eq4.1}
\end{equation}
in the entropy
\begin{equation}
S (M_0 , R_0) = c \, (R_0) \, a_0 \, M_0 \, , 
\label{eq4.2}
\end{equation}
is near unity. [We use Eq.~(\ref{eq2.30}) in the limit $g \rightarrow 0$, for which it was 
derived.] Because of this reduced sensitivity of $c \,(R_0)$ on a possible direct effect 
of $g$ on $R$ ({\it i.e.} $R(g) = R_0 + \delta_g \, R$), the main effect of self-gravity 
on the entropy (considered as a function of the actual values $M$, $R$ when $g \not= 0$) 
will come from replacing $M_0$ as a function of $M$ and $R$. The mass-shift result 
(\ref{eq3.25}) gives $\delta \, M = M - M_0$ to first order in $g^2$. To the same 
accuracy\footnote{Actually, Eq.~(\ref{eq4.3}) is probably a more 
accurate version of the mass-shift formula because it exhibits the real mass $M$ (rather 
than the bare mass $M_0$) as the source of self-gravity.}, (\ref{eq3.25}) gives $M_0$ as a 
function of $M$ and $R$:
\begin{equation}
M_0 \simeq M + c_3 \, g^2 \, \frac{M^2}{R^{d-2}} = M \left( 1 + c_3 \, \frac{g^2 \, 
M}{R^{d-2}} \right) \, , 
\label{eq4.3}
\end{equation}
where $c_3$ is a positive numerical constant.

Finally, combining Eqs.~(\ref{eq4.1})--(\ref{eq4.3}) (and neglecting, as just said, a 
small effect linked to $\delta_g \, R \not= 0$) leads to the following {\it relation 
between the entropy, the mass and the size (all considered for self-gravitating states, 
with $g \not= 0$)}
\begin{equation}
S(M,R) \simeq a_0 \, M \left( 1 - \frac{1}{R^2} \right) \left( 1 - \frac{R^2}{M^2} \right) 
\left( 1 + \frac{g^2 \, M}{R^{d-2}} \right) \, . 
\label{eq4.4}
\end{equation}
For notational simplicity, we henceforth set to unity (by using some redefinitions
and working in some suitably defined ``string units'')) 
the coefficients $c_1$, $c_2$ and $c_3$. The possibility of smoothly transforming 
self-gravitating string states into black-hole states come from the peculiar radius 
dependence of the entropy $S(M,R)$. Eq.~(\ref{eq4.3}) exhibits two effects varying in 
opposite directions: (i) self-gravity favors small values of $R$ (because they correspond 
to larger values of $M_0$, {\it i.e.} of the ``bare'' entropy), and (ii) the constraint of 
being of some fixed size $R$ disfavors both small $(R \ll \sqrt{M})$ and 
large $(R \gg \sqrt{M})$ values of $R$. For given values of $M$ and $g$, the most numerous 
(and therefore most probable) string states will have a size $R_* (M;g)$ which maximizes 
the entropy $S(M,R)$. Said differently, the total degeneracy of the complete ensemble of 
self-gravitating string states with total energy $M$ (and no {\it a 
priori} size restriction) will be given by an integral (where $\Delta \, R$ is the rms 
fluctuation of $R$)
\begin{equation}
{\cal D} (M) \sim \int \frac{d R}{\Delta \, R} \, e^{S(M,R)} \sim e^{S(M,R_*)} 
\label{eq4.5}
\end{equation}
which will be dominated by the saddle point $R_*$ which maximizes the exponent.

The value of the most probable size $R_*$ is a function of $M$, $g$ and the space 
dimension $d$. We refer to Ref.~\cite{DV00} for a full treatment. Let us only indicate the 
results in the (actual) {\it three-dimensional} case, $d=3$.
 When maximizing the entropy $S(M,R)$ with 
respect to $R$ one finds that: (i) when $g^2 \ll M^{-3/2}$, the most probable size $R_* 
(M,g) \sim \sqrt{M}$, (ii) when $g^2 \gg M^{-3/2}$,
\begin{equation}
R_* (M,g) \simeq \frac{1 + \sqrt{1+3 \lambda^2}}{\lambda} 
\label{eq4.1new}
\end{equation}
where $\lambda \equiv g^2 M$.

Eq.~(\ref{eq4.1new}) says that, when $g^2$ increases, and therefore when $\lambda$ 
increases (beyond $M^{-1/2}$) the typical size of a self-gravitating string decreases, and 
(formally) tends to a limiting size of order unity, $R_{\infty} = \sqrt 3$ ({\it i.e.} of 
order the string length scale $\ell_s = \sqrt{2 \alpha'}$) when $\lambda \gg 1$. However, 
the fractional self-gravity $G M / R_* \simeq \lambda / R_*$ (which measures 
the gravitational deformation away from flat space) becomes unity for $\lambda = \sqrt 5$ 
and formally increases without limit when $\lambda$ further increases. Therefore, we 
expect that for some value of $\lambda$ of order unity, the self-gravity of the compact 
string state already reached when $\lambda \sim 1$ (indeed, Eq.~(\ref{eq4.1new}) predicts 
$R_* \sim 1$ when $\lambda \sim 1$) will become so strong that it will (continuously) 
turn into a black-hole state. Having argued that the dynamical threshold for the 
transition string $\rightarrow$ black hole is $\lambda \sim 1$, we now notice that, for 
such a value of $\lambda$ the entropy $S(M) = S (M,R_* (M)) \simeq a_0 \, M \left[ 1 + 
\frac{1}{4} \, (g^2 M)^2 \right]$ of the string state (of mass $M$) {\it matches the 
Bekenstein-Hawking entropy} $S_{\rm BH} (M) \sim g^2 \, M^2 = \lambda M$ of the formed 
black 
hole. One further checks that the other global physical characteristics of the string 
state (radius $R_*$, luminosity $P$, Eq.~(\ref{eq3.29})) match those of a Schwarzschild 
black hole of the same mass ($R_{\rm BH} \sim GM \sim g^2 M \, , \, P_{\rm Hawking} \sim 
R_{\rm BH}^{-2}$) when $\lambda \sim 1$.

\begin{figure}[ht]
\begin{center}
\includegraphics[scale=.8]{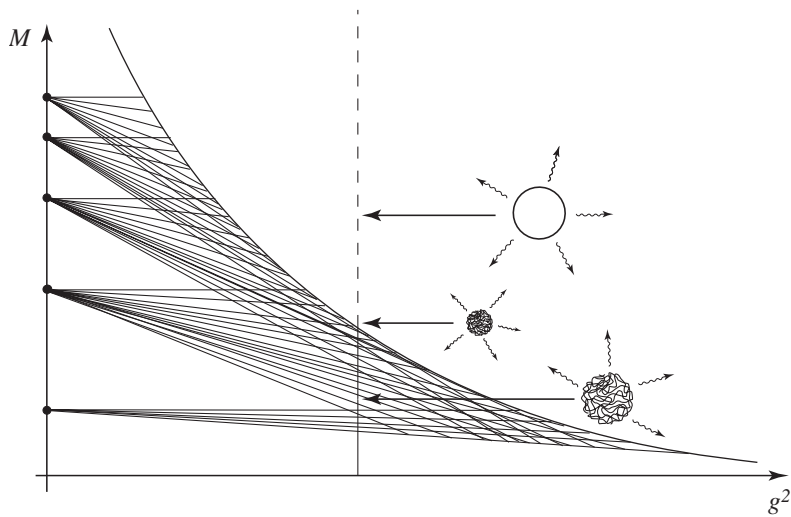}
\vglue 5mm
\caption{Evolution of the energy levels of massive string states as the
string coupling $g$ increases. String states are expected to transform into
black-hole states on the correspondence line $g^2 M \sim M_s$.
Vertical line: Corresponding transformation of a
black hole, loosing mass at fixed $g$ under Hawking radiation, into an initially compact 
string
state, which inflates before decaying into massless radiation. }
\label{fig4}
\end{center}
\end{figure}

Conceptually, the main new result of the analysis summarized above concerns the most
probable state of a very massive single\footnote{We consider states of a single string 
because, for large values of the mass, the single-string entropy approximates the total 
entropy up to subleading terms.} self-gravitating string.  By combining our estimates of 
the entropy reduction due to the size constraint, and of the mass shift we come up with 
the expression (\ref{eq4.4}) for the logarithm of the number of self-gravitating string 
states of size $R$.  Our analysis of the function $S(M,R)$ clarifies the 
correspondence~\cite{S93,halyo,HP97,HP98} between string states and black holes.  In 
particular, our results confirm many of the results of~\cite{HP98}, but make them (in our 
opinion) physically clearer by dealing directly with the size distribution, in real space, 
of an ensemble of string states.  When our results differ from those of~\cite{HP98}, they 
do so in a way which simplifies the physical picture and make even more compelling the 
existence of a correspondence between strings and black holes. The simple physical picture 
suggested\footnote{Our conclusions are not rigourously established because they rely on 
assuming the validity of the result (\ref{eq4.4}) beyond the domain ($R^{-2} \ll 1$, $g^2 
\, M / R^{d-2} \ll 1$) where it was derived. However, we find heuristically convincing to 
believe in the presence of a reduction factor of the type $1-R^{-2}$ down to sizes very 
near the string scale. Our heuristic dealing with self-gravity is less compelling because 
we do not have a clear signal of when strong gravitational field effects become 
essential.} by our results is the following: In any dimension, if we start with a massive 
string state and increase the string coupling $g$, a {\it typical} string state will,
eventually, become more compact and will end up, when $\lambda_c = g_c^2\, M \sim 1$, in a 
``condensed state'' of size $R \sim 1$, and mass density $\rho \sim g_c^{-2}$. Note that 
the basic reason why small strings, $R \sim 1$, dominate in any dimension the entropy when 
$\lambda \sim 1$ is that they descend from string states with bare mass $M_0 \simeq M (1 + 
\lambda / R^{d-2}) \sim 2 M$ which are exponentially more numerous than less condensed 
string states corresponding to smaller bare masses (see Fig.~\ref{fig4}).

The nature of the transition between the initial ``dilute'' state and the final 
``condensed'' one depends on the value of the space dimension $d$. In $d=3$, the transition 
is gradual: when $\lambda < M^{-1/2}$ the size of a typical state is $R_*^{(d=3)} \simeq 
M^{1/2}(1-M^{1/2} \, \lambda / 8)$, when $\lambda >  M^{1/2}$ the typical size is 
$R_*^{(d=3)} \simeq (1 + (1+3 \, \lambda^2)^{1/2}) / \lambda$. In $d=4$, the transition 
toward a condensed state is still continuous, but most of the size evolution takes place 
very near $\lambda = 1$: when $\lambda <1$, $R_*^{(d=4)} \simeq M^{1/2} 
(1-\lambda)^{1/4}$, and when $\lambda > 1$,$R_*^{(d=4)} \simeq (2\lambda / (\lambda - 
1))^{1/2}$, with some smooth blending between the two evolutions around $\vert \lambda - 1 
\vert \sim M^{-2/3}$. In $d \geq 5$, the transition is discontinuous (like a first order 
phase transition between, say, gas and liquid states). Barring the consideration of 
metastable (supercooled) states, on expects that when $\lambda = \lambda_2 \simeq 
\nu^{\nu} / (\nu - 1)^{\nu - 1}$ (with $\nu = (d-2) /2$), the most probable size of a 
string state will jump from $R_{\rm rw}$ (when $\lambda < \lambda_2$) to a size of order 
unity (when $\lambda > \lambda_2$).

One can think of the ``condensed'' state of (single) string matter, reached (in any $d$) 
when $\lambda \sim 1$, as an analog of a neutron star with respect to an ordinary star (or 
a white dwarf). It is very compact (because of self gravity) but it is stable (in some 
range for $g$) under gravitational collapse. However, if one further increases $g$ or $M$ 
(in fact, $\lambda = g^2 \, M$), the condensed string state is expected (when $\lambda$ 
reaches some $\lambda_3 > \lambda_2$, $\lambda_3 = {\cal O} (1)$) to collapse down to a 
black-hole state (analogously to a neutron star collapsing to a black hole when its mass 
exceeds the Landau-Oppenheimer-Volkoff critical mass). Still in analogy with neutron 
stars, one notes that general relativistic strong gravitational field effects are crucial 
for determining the onset of gravitational collapse; indeed, under the ``Newtonian'' 
approximation (\ref{eq4.4}), the condensed string state could continue to exist for 
arbitrary large values of $\lambda$.

It is interesting to note that the value of the mass density at the formation of the 
condensed string state is $\rho \sim g^{-2}$. This is reminiscent of the prediction by 
Atick and Witten~\cite{AW88} of a first-order phase transition of a self-gravitating 
thermal gas of strings, near the Hagedorn temperature\footnote{Note that, by definition, 
in our {\it single} string system, the formal temperature $T = (\partial S / \partial 
M)^{-1}$ is always near the Hagedorn temperature.}, towards a dense state with energy
density $\rho \sim g^{-2}$ (typical of a genus-zero contribution to the free energy). 
Ref.~\cite{AW88} suggested that this transition is first-order because of the coupling to 
the dilaton. This suggestion agrees with our finding of a discontinuous transition to the 
single string condensed state in dimensions $\geq 5$ (Ref.~\cite{AW88} works in higher 
dimensions, $d=25$ for the bosonic case). It would be interesting to deepen these links 
between self-gravitating single string states and multi-string states.

Let us come back to the consequences of the picture summarized above for the problem of 
the end point of the evaporation of a Schwarzschild black hole and the interpretation of 
black-hole entropy. See Fig.~\ref{fig4}.

 In that case one fixes the value of $g$ (assumed to be $\ll 1$) and 
considers a black hole which slowly looses its mass via Hawking radiation. When the mass 
gets as low as a value\footnote{Note that the mass at the black hole $\rightarrow$ 
string transition is larger than the Planck mass $M_P \sim (G)^{-1/2} \sim g^{-1}$ by a 
factor $g^{-1} \gg 1$.} $M \sim g^{-2}$, for which the radius of the black hole is of 
order one (in string units), one expects the black hole to transform (in all dimensions) 
into a typical string\footnote{A check on the single-string dominance of the transition 
black hole $\rightarrow$ string is to note that the single string entropy $\sim M / M_s$ 
is much larger than the entropy of a ball of radiation $S_{\rm rad} \sim (RM)^{d/(d+1)}$ 
with size $R \sim R_{\rm BH} \sim \ell_s$ at the transition.} state corresponding to 
$\lambda = g^2 \, M \sim 1$, which is a dense state (still of radius $R \sim 1$). This 
string state will further decay and loose mass, predominantly via the emission of massless 
quanta, with a quasi thermal spectrum with temperature $T \sim T_{\rm Hagedorn} = 
a_0^{-1}$, with $a_0$  defined after (\ref{eq2.30}), and  which smoothly matches the 
previous black 
hole Hawking temperature. This mass 
loss will further decrease the product $\lambda = g^2 \, M$, and this decrease will, either 
gradually or suddenly, cause the initially compact string state to inflate to much larger 
sizes. For instance, if $d \geq 4$, the string state will quickly inflate to a size $R 
\sim \sqrt{M}$. Later, with continued mass loss, the string size will slowly shrink again 
toward $R \sim 1$ until a remaining string of mass $M \sim 1$ finally decays into stable 
massless quanta. In this picture, the black-hole entropy acquires a somewhat clear 
statistical significance (as the degeneracy of a corresponding typical string state) {\it 
only} when $M$ and $g$ are related by $g^2 \, M \sim 1$. If we allow ourselves to vary (in 
a Gedanken experiment) the value of $g$ this gives a potential statistical significance to 
any black-hole entropy value $S_{\rm BH}$ (by choosing $g^2 \sim S_{\rm BH}^{-1}$). We do 
not claim, however, to have a clear idea of the direct statistical meaning of $S_{\rm BH}$ 
when $g^2 \, S_{\rm BH} \gg 1$. Neither do we clearly understand the fate of the very large 
space (which could be excited in many ways) which resides inside very large classical 
black holes of radius $R_{\rm BH}\sim (g^2 \, S_{\rm BH})^{1/(d-1)} \gg 1$. The fact that 
the interior of a black hole of given mass could be {\it arbitrarily} large\footnote{E.g., 
in the Oppenheimer-Snyder model, one can join an arbitrarily large closed Friedmann dust 
universe, with hyperspherical opening angle $0 \leq \chi_0 \leq \pi$ arbitrarily near 
$\pi$, onto an exterior Schwarzschild spacetime of given mass $M$.}, and therefore 
arbitrarily complex, suggests that black-hole physics is not exhausted by the idea 
(discussed here) of a reversible transition between string-length-size black holes and 
string states.

On the string side, we also do not clearly understand how one could follow in detail (in 
the present non BPS framework) the ``transformation'' of a strongly self-gravitating 
string state into a black-hole state.

Finally, let us note that we expect that self-gravity will lift nearly completely the 
degeneracy of string states. [The degeneracy linked to the rotational symmetry, {\it i.e.} 
$2J + 1$ in $d=3$, is probably the only one to remain, and it is negligible compared to 
the string entropy.] Therefore we expect (contrary to what is assumed in many works 
\cite{B74,M86,K86,M94,B97,Hod,Dreyer}) that the separation $\delta \, E$ between 
subsequent (string and black-hole) energy levels will be exponentially small: $\delta \, E 
\sim \Delta \, M \, \exp (-S(M))$, where $\Delta \, M$ is the canonical-ensemble 
fluctuation in $M$. 
See Fig.~\ref{fig4} which sketches how self-gravity (which depends on the ``size''
of the considered massive string state) lifts the degeneracy of free string states,
and is expected to lead to a very high density of (nearly non-degenerate)
energy levels. [This picture of the level structure of non-supersymmetric
string/black-hole states should be contrasted with the case of supersymmetric
BPS $D$-brane/black-hole states, which features well-separated discrete levels
having exponentially large degeneracies. See Fig.~\ref{fig3} and Eq.~(\ref{Dmass}).]
Such a $\delta \, E$ is negligibly small compared to the 
radiative width $\Gamma \sim g^2 \, M$ of the levels. This seems to mean that the 
discreteness of the quantum levels of strongly self-gravitating strings and black holes is 
very much blurred, and difficult to see observationally. Such exponentially small $\delta 
E$
is also expected to play an important role in ``explaining'' why sufficiently massive 
string
states behave as a thermal state (see, in this respect, the recent discussion \cite{DRS02}
of the ``self-thermalization'' of pure states).

\section{Conclusions}

We hope that this short review has allowed the reader to grasp the richness of the issue 
of black-hole entropy. Thirty years after the original proposal of Bekenstein and the 
discovery by Hawking of the quantum radiance of black holes, the issue remains somewhat 
cloudy. String theory has brought very significant progress by allowing one to interpret 
the microscopic degrees of freedom of certain {\it extremal} (supersymmetric) black holes 
in terms of a sort of ``analytic continuation in the string coupling  $g$'' of systems 
of very massive, (Ramond-Ramond) charged string solitons (Dirichlet branes). However, it 
is frustrating that the $D$-brane description does not apply to the regime of large 
(effective) string coupling $g$ where these systems are believed to ``transform into'' 
black holes. We have sketched the limited progress made in string theory for understanding 
the microscopic degrees of freedom of {\it non extremal} (non supersymmetric) black holes, 
and, in particular, the simplest Schwarzschild ones. A plausible picture emerged of a 
reversible transformation between Schwarzschild black-hole states (when $g^2 M \gtrsim 
M_s$) into self-gravitating massive string states (when $g^2 M \lesssim M_s$). However, 
the lack of supersymmetry for such states limits the secure applicability of present 
string technology to the regime $g^2 M \ll M_s$. The extension of the results to $g^2 M 
\sim M_s$ is physically plausible but not technically justified. In addition, all stringy 
calculations fail to give a precise Hilbert-space description of the microscopic structure 
of black holes in the regime ($g^2 M \gg M_s$, or its BPS equivalent) where the considered 
``internal structure'' has become a large, quasi-classical black hole. In particular, the 
quantum description of the large spacetime region constituting the interior of a 
quasi-classical black hole (see Fig.~\ref{fig1}), and of all the excitations it contains 
(such as the ``lost'' parts of the Hawking particle creation phenomenon), remain somewhat 
mysterious (see \cite{HM03} for a recent attempt at describing the quantum state of a 
quasi-classical
black hole in a unitary way). Clearly the issue of black-hole entropy will continue to 
exercize the sagacity 
of many researchers for many years to come, and will continue to provide one of our few 
handles on the physics of quantum relativistic gravity, {\it i.e.} the domain where 
$\hbar$, $c$ and $G$ all play important roles.

\section*{Acknowledgements}

I wish to thank Bertrand Duplantier for useful comments, Vincent Pasquier for bringing 
Wheeler's 
recollection to my attention, and the Kavli Institute for 
Theoretical Physics for hospitality
while this review was completed.  This work was supported in part by the
National Science Foundation under Grant No. PHY99-07949.

\section*{Appendix}
Here is an excerpt from John Wheeler's book \cite{wheeler}
 about the genesis of the discovery of
the concept of black-hole entropy by Jacob Bekenstein:

\noindent``One afternoon in 1970 Bekenstein--then a graduate student--and I were
discussing black-hole physics in my office in Princeton's Jadwin Hall.
I told him the concern I always feel when a hot cup of tea exchanges heat energy 
with a cold cup of tea.
By allowing that transfer of heat I do not alter the energy of the universe, but 
I do increase its microscopic disorder, its information loss, its entropy.
The entropy of the world always increases in an irreversible process like that.

{\it ``The consequence of my crime, Jacob, echo down to the end of time,''}
I noted. {\it ``But if a black hole swims by, and I drop the teacups into it, I
conceal from all the world the evidence of my crime. How remarkable!''}

Bekenstein, a man of deep integrity, takes the lawfulness of creation as a 
matter of utmost seriousness. Several months later, he came back with a 
remarkable idea. {\it ``You don't destroy entropy when you drop those teacups into the 
black hole. The black hole already has entropy, and you only increase it!''}

Bekenstein went on to explain that the surface area of a black hole is not
only analogous to entropy, it is entropy, and the surface gravity of a black
hole (measured, for example, by the downward acceleration of a rock as it
crosses the horizon) is not only analogous to temperature, it is temperature.
A black hole is not totally cold!...''

\end{document}